\documentstyle[preprint,eqsecnum,aps]{revtex}
\begin{document}



\def\Rnum{{\bf R}}
\def\Cnum{{\bf C}}

\def\eqref#1{(\ref{#1})}
\def\eqrefs#1#2{(\ref{#1}) and~(\ref{#2})}
\def\eqsref#1#2{(\ref{#1}) to~(\ref{#2})}

\def\Eqref#1{Eq.~(\ref{#1})}
\def\Eqrefs#1#2{Eqs.~(\ref{#1}) and~(\ref{#2})}
\def\Eqsref#1#2{Eqs.~(\ref{#1}) to~(\ref{#2})}

\def\secref#1{Sec.~\ref{#1}}
\def\secrefs#1#2{Secs.~\ref{#1} and~\ref{#2}}
\def\secsref#1#2{Secs.~\ref{#1} to~\ref{#2}}

\def\appref#1{App.~\ref{#1}}

\def\Ref#1{Ref.\cite{#1}}
\def\Refs#1{Refs.\cite{#1}}

\def\Cite#1{${\mathstrut}^{\cite{#1}}$}

\def\tableref#1{Table~\ref{#1}}

\def\figref#1{Fig.~\ref{#1}}

\hyphenation{Eq Eqs Sec App Ref Fig}

\def\demo #1. #2\par{\medbreak\noindent{\bf#1.\enspace}{\rm#2}\par\medbreak}

\def\EQ{\begin{equation}}
\def\EQs{\begin{eqnarray}}
\def\endEQ{\end{equation}}
\def\endEQs{\end{eqnarray}}

\def\eqtext#1{\hbox{\rm{#1}}}

\def\proclaim#1{\medbreak
\noindent{\it {#1}}\par\medbreak}
\def\Proclaim#1#2{\medbreak
\noindent{\bf {#1}}{\it {#2}}\par\medbreak}


\def\fewquad{\qquad\qquad}
\def\severalquad{\qquad\fewquad}
\def\manyquad{\qquad\severalquad}
\def\manymanyquad{\manyquad\manyquad}

\def\sub#1{
\setbox1=\hbox{{$\scriptscriptstyle #1$}} 
\dimen1=0.6\ht1
\mkern-2mu \lower\dimen1\box1 \hbox to\dimen1{\box1\hfill} }

\def\eqtext#1{\hbox{\rm{#1}}}

\def\endproof{
\setbox2=\hbox{{$\sqcup$}} \setbox1=\hbox{{$\sqcap$}} 
\dimen1=\wd1
\box2\kern-\dimen1 \hbox to\dimen1{\box1} }

\def\mstrut{\mathstrut}
\def\hp#1{\hphantom{#1}}

\def\mixedindices#1#2{{\mstrut}^{\mstrut #1}_{\mstrut #2}}
\def\downindex#1{{\mstrut}^{\mstrut}_{\mstrut #1}}
\def\upindex#1{{\mstrut}_{\mstrut}^{\mstrut #1}}
\def\downupindices#1#2{{\mstrut}_{\mstrut #1}^{\hp{#1}\mstrut #2}}
\def\updownindices#1#2{{\mstrut}^{\mstrut #1}_{\hp{#1}\mstrut #2}}

\def\index#1{{\scriptstyle #1}}

\def\Parder#1#2{
\mathchoice{\partial{#1} \over\partial{#2}}{\partial{#1}/\partial{#2}}{}{} }
\def\parder#1{\partial/\partial{#1}}

\def\der#1{\partial\downindex{#1}}
\def\coder#1{\partial\upindex{#1}}
\def\D#1{D\downindex{#1}}
\def\coD#1{D\upindex{#1}}
\def\covder#1{\nabla\downindex{#1}}
\def\covcoder#1{\nabla\upindex{#1}}
\def\g#1{g\downindex{#1}}
\def\flat#1{\eta\downindex{#1}}
\def\invflat#1{\eta\upindex{#1}}
\def\vol#1#2{\epsilon\downupindices{#1}{#2}}
\def\e#1#2{e\downupindices{#1}{#2}}
\def\inve#1#2{e\updownindices{#1}{#2}}
\def\x#1#2{x\mixedindices{#1}{#2}}
\def\id#1#2{\delta\mixedindices{#2}{#1}}

\def\Q#1#2{Q\downupindices{#2}{#1}}
\def\tQ#1#2{{\widetilde Q}\downupindices{#2}{#1}}
\def\cQ#1#2{{\bar Q}\downupindices{#2}{#1}}
\def\bigQ{{\bf Q}}
\def\X#1{X\downindex{#1}}
\def\tX#1{{\tilde X}\downindex{#1}}
\def\bigX{{\bf X}}

\def\XQ#1{\chi\upindex{#1}}
\def\tXQ#1{{\tilde\chi}\upindex{#1}}

\def\curr#1#2#3#4{\Psi^{#1}_{\rm #2}\mixedindices{#3}{#4}}
\def\dens#1{\Psi\upindex{#1}}
\def\denshat#1{{\hat\Psi}\upindex{#1}}
\def\H#1#2{\Phi\mixedindices{#1}{#2}}
\def\TH#1#2#3{\Phi^{(#1)}_T\mixedindices{#2}{#3}}
\def\ZH#1#2#3{\Phi^{(#1)}_Z\mixedindices{#2}{#3}}
\def\VH#1#2#3{\Phi^{(#1)}_V\mixedindices{#2}{#3}}
\def\WH#1#2{\Phi_W\mixedindices{#1}{#2}}

\def\curl#1#2{\Theta\mixedindices{#1}{#2}}
\def\tcurl#1#2{\tilde\Theta\mixedindices{#1}{#2}}
\def\triv#1#2{\Upsilon\mixedindices{#1}{#2}}

\def\P#1#2{P\downupindices{#2}{#1}}
\def\tP#1#2{{\widetilde P}\downupindices{#2}{#1}}
\def\bigP{{\bf P}}

\def\zeroP#1#2{{U}\downupindices{#2}{#1}}
\def\firstP#1#2{{V}\downupindices{#2}{#1}}
\def\zerocP#1#2{{\bar U}\downupindices{#2}{#1}}
\def\firstcP#1#2{{\bar V}\downupindices{#2}{#1}}
\def\W#1#2{W\downupindices{#2}{#1}}
\def\cW#1#2{\bar W\downupindices{#2}{#1}}

\def\TQ#1#2#3{Q_T^{(#1)}\downupindices{#3}{#2}}
\def\ZQ#1#2#3{Q_Z^{(#1)}\downupindices{#3}{#2}}
\def\VQ#1#2#3{Q_V^{(#1)}\downupindices{#3}{#2}}
\def\WQ#1#2{Q_W\downupindices{#2}{#1}}

\def\sF#1#2{\phi\mixedindices{#1}{#2}}
\def\csF#1#2{{\bar\phi}\mixedindices{#1}{#2}}
\def\sFder#1#2#3#4{\phi\mixedindices{#1\hp{#2}#3}{\hp{#1}#2#4}}
\def\csFder#1#2#3#4{{\bar\phi}\mixedindices{#1\hp{#2}#3}{\hp{#1}#2#4}}
\def\snF#1#2#3#4{\phi\mixedindices{(#1)}{#2}\mixedindices{#3}{#4}}
\def\csnF#1#2#3#4{\bar\phi\mixedindices{(#1)}{#2}\mixedindices{#3}{#4}}

\def\sder#1#2{\partial\mixedindices{#1}{#2}}
\def\covsder#1#2{\nabla\mixedindices{#1}{#2}}
\def\sD#1#2{D\mixedindices{#1}{#2}}

\def\sxparder#1#2{\parder{\x{#1}{#2}}}
\def\sFparder#1#2#3#4{\parder{\sFder{#1}{#2}{#3}{#4}}}
\def\csFparder#1#2#3#4{\parder{\csFder{#1}{#2}{#3}{#4}}}

\def\sFsol#1#2{\varphi\mixedindices{#1}{#2}}
\def\csFsol#1#2{{\bar\varphi}\mixedindices{#1}{#2}}
\def\sFdersol#1#2#3#4{\varphi\mixedindices{#1\hp{#2}#3}{\hp{#1}#2#4}}
\def\csFdersol#1#2#3#4{{\bar\varphi}\mixedindices{#1\hp{#2}#3}{\hp{#1}#2#4}}

\def\spinor#1#2#3{{#1}\mixedindices{#2}{#3}}
\def\cspinor#1#2#3{\bar{#1}\mixedindices{#2}{#3}}

\def\FE#1#2{\Delta\mixedindices{#1}{#2}}
\def\cFE#1#2{\bar\Delta\mixedindices{#1}{#2}}

\def\KV#1#2{\xi\mixedindices{#1}{#2}}
\def\othKV#1#2{\zeta\mixedindices{#1}{#2}}
\def\prodKV#1#2#3{\varrho^{#1}\mixedindices{#2}{#3}}
\def\cKV#1#2{{\bar\xi}\mixedindices{#1}{#2}}
\def\cothKV#1#2{\bar\zeta\mixedindices{#1}{#2}}
\def\KS#1#2{\kappa\mixedindices{#1}{#2}}
\def\cKS#1#2{{\bar\kappa}\mixedindices{#1}{#2}}
\def\K#1#2{K\updownindices{#1}{#2}}
\def\KY#1#2{Y\updownindices{#1}{#2}}
\def\othKY#1#2{\Upsilon\updownindices{#1}{#2}}
\def\prodKY#1#2#3{Y^{#1}\updownindices{#2}{#3}}
\def\cKY#1#2{\bar Y\updownindices{#1}{#2}}
\def\cothKY#1#2{\bar\Upsilon\updownindices{#1}{#2}}
\def\w#1#2{\omega\downupindices{#2}{#1}}
\def\cw#1#2{\bar\omega\downupindices{#2}{#1}}

\def\so#1#2{o\mixedindices{#1}{#2}}
\def\si#1#2{\iota\mixedindices{#1}{#2}}
\def\cso#1#2{{\bar o}\mixedindices{#1}{#2}}
\def\csi#1#2{{\bar\iota}\mixedindices{#1}{#2}}
\def\l#1{\ell\upindex{#1}}
\def\n#1{n\upindex{#1}}
\def\m#1{m\upindex{#1}}
\def\cm#1{\bar m\upindex{#1}}

\def\a#1#2{\alpha\mixedindices{#1}{#2}}
\def\b#1#2{\beta\mixedindices{#2}{#1}}
\def\ca#1#2{\bar\alpha\mixedindices{#1}{#2}}
\def\cb#1#2{\bar\beta\mixedindices{#2}{#1}}
\def\ta#1{\tilde\alpha\upindex{#1}}
\def\tb#1{\tilde\beta\downindex{#1}}

\def\t#1#2{t\downupindices{#2}{#1}}
\def\r#1#2{r\mixedindices{#1}{#2}}
\def\u#1#2{u\downupindices{#2}{#1}}
\def\z#1#2{n\downupindices{#2}{#1}}

\def\S#1#2{S\mixedindices{#1}{#2}}
\def\T#1#2#3{{#1}\mixedindices{#2}{#3}}

\def\E#1#2{\vec E\mixedindices{#1}{#2}}
\def\B#1#2{\vec B\mixedindices{#1}{#2}}
\def\cE#1#2{\vec{\bar E}\mixedindices{#1}{#2}}
\def\vecD#1{\vec D\downindex{#1}}
\def\veccoD#1{\vec D\upindex{#1}}

\def\Lie#1{{\cal L}_{#1}}
\def\sLie#1{\hat{\cal L}_{#1}}
\def\csLie#1{\hat{\cal L}_{#1}}
\def\div{{\rm div\,}}

\def\C#1#2{{{#1}\choose{#2}}}
\def\Cquant#1#2{C^{#1}_{#2}}

\def\vs#1#2{{\cal V}\mixedindices{#2}{#1}}
\def\ks#1#2{{\cal K}\mixedindices{#2}{#1}}
\def\qvs#1#2{{\cal N}\mixedindices{#2}{#1}}

\def\i{{\rm i}}

\def\const{{\rm const}}

\def\frac#1#2{{#1\over #2}}

\def\Kvec/{Killing vector}
\def\Kspin/{Killing spinor}
\def\Kten/{Killing tensor}
\def\KYten/{Killing-Yano tensor}

\def\ie/{i.e.}
\def\eg/{e.g.}

\hyphenation{
}

\narrowtext
\tightenlines

\title{ Conserved currents of massless fields of spin $s\ge 1/2$ }

\author{ Stephen C. Anco${}^1$\cite{email1}
and Juha Pohjanpelto${}^2$\cite{email2} }

\address{
${}^1$Department of Mathematics\\
Brock University,
St. Catharines, ON L2S 3A1, Canada\\
${}^2$Department of Mathematics\\
Oregon State University,
Corvallis, OR 97331-4605, U.S.A }

\maketitle

\begin{abstract}
A complete and explicit classification of
all locally constructed conserved currents
and underlying conserved tensors
is obtained for
massless linear symmetric spinor fields of any spin $s\ge 1/2$
in four dimensional flat spacetime.
These results generalize the recent classification
in the spin $s=1$ case of all conserved currents
locally constructed from the electromagnetic spinor field.
The present classification yields spin $s\ge 1/2$ analogs of
the well-known electromagnetic stress-energy tensor
and Lipkin's zilch tensor,
as well as a spin $s\ge 1/2$ analog of
a novel chiral tensor found in the spin $s=1$ case.
The chiral tensor possesses odd parity under a duality symmetry
(\ie/, a phase rotation) on the spin $s$ field,
in contrast to the even parity of the stress-energy and zilch tensors.
As a main result, it is shown that 
every locally constructed conserved current for each $s\ge 1/2$ 
is equivalent to a sum of
elementary linear conserved currents,
quadratic conserved currents associated with
the stress-energy, zilch, and chiral tensors,
and higher derivative extensions of these currents
in which the spin $s$ field is replaced by
its repeated conformally-weighted Lie derivatives
with respect to conformal Killing vectors of flat spacetime.
Moreover, all of the currents have a direct, unified characterization
in terms of \Kspin/s.
The cases $s=2$, $s=1/2$ and $s=3/2$ provide a complete set of
conserved quantities for propagation of gravitons
(\ie/, linearized gravity waves),
neutrinos and gravitinos, respectively,
on flat spacetime.
The physical meaning of
the zilch and chiral quantities is discussed.
\end{abstract}
\newpage

\section{Introduction}

In classical relativity theory,
the fundamental spinor equations describing propagation of
free massless spin $s\ge 1/2$ fields in spacetime
have many interesting aspects
\cite{Penrose}.
In particular, due to their linear nature, these field equations
have long been known to admit a rich structure of
conserved currents and symmetries,
although few complete results on this structure
have been obtained to-date
\cite{FushchichNikitin-book}. 

A main motivation for studying conserved currents of
massless spin $s$ fields is to find
a complete set of conserved quantities
characterizing the propagation of electromagnetic waves
and linearized gravity waves in the cases $s=1,2$,
as well as the propagation of neutrinos in the case $s=1/2$
(and more theoretically, gravitinos in the case $s=3/2$).
In recent work \cite{maxwell,maxwellsymm}
on the spin $1$ field equations
(\ie/, Maxwell's equations) in flat spacetime,
using spinorial techniques and general conservation law methods
\cite{Olver,AncoBluman1,AncoBlumanII},
we obtained a complete and explicit classification of
all conserved currents that are locally constructed from the spin $1$
field strength and its derivatives to any order.
A principal result for the spin $1$ field equations 
is that in addition to
the elementary linear currents
and the well-known quadratic stress-energy and zilch currents,
there are also 
quadratic chiral currents \cite{FushchichNikitin-book,maxwell}
which, in contrast to the stress-energy and zilch currents,
possess odd parity under the duality symmetry
interchanging the electric and magnetic
components of the spin $1$ field strength.
Moreover, all of these currents have higher derivative extensions \cite{survey}
obtained by a repeated replacement of the field strength
by conformally-weighted Lie derivatives with respect to conformal \Kvec/s
of flat spacetime.
Our classification establishes that every conserved current
locally constructed from the spin $1$ field strength and its derivatives
is equivalent to a linear combination of
the elementary currents, 
the even parity stress-energy and zilch currents,
the odd parity chiral currents, 
and their higher derivative extensions.

In this paper we generalize the previous classification results
on conserved currents
to the massless spin $s$ field equations for all $s=1/2,1,3/2,2,\ldots$
in flat spacetime.
Besides spin $s$ analogs of
the electromagnetic stress-energy and zilch conserved currents,
we also obtain new chiral conserved currents
analogous to those in the spin $1$ case,
as well as higher derivative extensions of these currents
given by a repeated application of
conformally-weighted Lie derivatives on the spin $s$ field.
We show that the resulting set of quadratic currents,
together with the set of elementary linear conserved currents,
yields a complete and unified classification of
all conserved currents
locally constructed from the spin $s$ field strength and its derivatives
to any order.
Furthermore, we derive underlying conserved tensors
associated with these conserved currents,
giving a spin $s$ generalization of the electromagnetic
energy tensor, zilch tensor, and new chiral tensor from \Ref{maxwell}.

In \secref{results} we state the main results
and outline our classification method,
which makes essential use of adjoint symmetries of
the massless spin $s$ field equations.
Adjoint symmetries are solutions of the formal adjoint equations of
the determining equations for symmetries,
and they give rise to conserved currents through a conservation law identity
involving the scaling symmetry of the spin $s$ field.
Hence, our analysis involves a complete classification of
the adjoint symmetries of
massless spin $s\ge 1/2$ fields in flat spacetime.
These are found to be characterized in terms of \Kspin/s,
which are spinorial generalizations of \Kvec/s related to twistors
\cite{Penrose}.
\Kspin/s in flat spacetime admit a factorization into
sums of symmetrized products of twistors.
This important property
allows the adjoint symmetries of massless spin $s$ fields
to be classified in a simple and uniform way,
which is pivotal for our results.
We carry out the classification analysis of adjoint symmetries
in \secref{adjointsymms}.

In \secref{currents} we present the details of
our classification of conserved currents.
The resulting set of conserved quantities obtained from these currents
is exhibited in \secref{quantities}
and some aspects of their physical meaning are discussed.
Finally, we make some concluding remarks in \secref{remarks}.

\section{Method and main results}
\label{results}

In four dimensional flat spacetime $M=(R^4,\flat{ab})$,
the spinor equations describing free massless spin $s\ge 1/2$ fields
are given by
\EQ
\sder{A_1}{A'} \sF{}{A_1\cdots A_{2s}} (x) =0 ,
\label{Feq}
\endEQ
where $\sF{}{A_1\cdots A_{2s}}$ is a symmetric spinor
representing the spin $s$ field strength
\cite{Penrose}.
Here $\sder{}{AA'}=\inve{a}{AA'}\der{a}$
is the spinorial derivative associated with
the metric compatible derivative $\der{a}$ on $M$,
\ie/, $\der{c}\flat{ab}=0$;
$\inve{a}{AA'}$ is the soldering form given by
a complex-valued null tetrad basis for the metric $\flat{ab}$
satisfying
$\inve{a}{AA'} \inve{b}{BB'} \flat{ab} = \vol{AB}{}\vol{A'B'}{}$,
where $\vol{AB}{}$ is the spin metric.
Throughout this paper
we use the index notation and conventions of \Ref{Penrose}; 
the metric signature is $(+,-,-,-)$,
and the spin metric is used for raising/lowering indices.
Note that, in standard Minkowski coordinates $\x{\mu}{}$,
the components of the derivative operators $\der{a}$ and $\sder{}{AA'}$
are simply the coordinate partial derivatives
$\der{\mu} = \parder{}{\x{\mu}{}}$
and $\inve{\mu}{AA'} \der{\mu} = \parder{}{\x{AA'}{}}$,
respectively,
with $\x{AA'}{}=\e{\mu}{AA'} \x{\mu}{}$.

It is convenient to introduce the jet spaces $J^q(\phi)$,
$0\leq q\leq\infty$,
using the spinorial coordinates
\EQ
J^q(\phi) =\{(
\x{CC'}{},\sF{}{A_1\cdots A_{2s}},\sFder{}{A_1\cdots A_{2s},}{C'_1}{C_1},
\ldots, \sFder{}{A_1\cdots A_{2s},}{C'_1\cdots C'_q}{C_1\cdots C_q} )\} ,
\label{Jpoint}
\endEQ
where the coordinates of a point in $J^q(\phi)$
are identified with a spacetime point $\x{CC'}{}$
and the values of the spin $s$ field strength
$\sF{}{A_1\cdots A_{2s}}(x)$
and its derivatives
$\sder{C'_1}{C_1}\cdots\sder{C'_p}{C_p}\sF{}{A_1\cdots A_{2s}}(x)$,
$1\le p\le q$, at $\x{CC'}{}$.
The solution space of the field equations \eqref{Feq}
is the subspace $R(\phi) \subset J^1(\phi)$
defined by imposing
$\sFder{}{A_1\cdots A_{2s},}{A_1}{A'}=0$.
The derivatives of the field equations \eqref{Feq} up to order $q$
similarly define the $q$-fold prolonged solution space
$R^q(\phi) \subset J^{q+1}(\phi)$
which is the subspace satisfying
$\sFder{}{A_1\cdots A_{2s},}{A_1C_1\cdots C_p}{A'C'_1\cdots C'_p}=0$,
$p\le q$.
As shown by Penrose \cite{Penrose},
the symmetric spinors
\EQ
\sFder{}{A_1\cdots A_{2s}}{C'_1\cdots C'_p}{C_1\cdots C_p} =
\sFder{}{(A_1\cdots A_{2s},}{(C'_1\cdots C'_p)}{\ C_1\cdots C_p)},
\quad
p\ge 0,
\label{symmdervar}
\endEQ
(where we let
$\sFder{}{A_1\cdots A_{2s},}{C'_1\cdots C'_p}{C_1\cdots C_p}$
for $p=0$ stand for $\sF{}{A_1\cdots A_{2s}}$)
form an ``exact set of fields''
that provide coordinates on the prolonged solution spaces,
\EQ
R^{q}(\phi) =\{(
\x{CC'}{},\sF{}{A_1\cdots A_{2s}},
\sFder{}{A_1\cdots A_{2s}}{C'_1}{C_1},\ldots,
\sFder{}{A_1\cdots A_{2s}}{C'_1\cdots C'_{q+1}}{C_1\cdots C_{q+1}} )\},
\quad
0\leq q\leq\infty .
\label{Rpoint}
\endEQ
Note that $R^\infty(\phi)$ is invariant under the transformation
\EQ
\sFder{}{A_1\cdots A_{2s},}{C'_1\cdots C'_p}{C_1\cdots C_p}
\rightarrow
-\i\sFder{}{A_1\cdots A_{2s},}{C'_1\cdots C'_p}{C_1\cdots C_p} ,
\quad p\ge 0 ,
\label{sduality}
\endEQ
which is called the duality symmetry of
the field equations \eqref{Feq}.

Write $\sD{}{AA'} = \inve{a}{AA'} \D{a}$
for the spinorial total derivative operator on $J^\infty(\phi)$
given by
\EQs
\sD{A'}{A} = \overline{ \sD{A}{A'} }
= &&
\sxparder{A}{A'} +\sum_{q\geq 0}
( \sFder{}{A_1\cdots A_{2s},}{A'C'_1\dots C'_q}{AC_1\dots C_q}
\sFparder{}{A_1\cdots A_{2s},}{C'_1\dots C'_q}{C_1\dots C_q}
\nonumber\\&&
+ \csFder{A'_1\cdots A'_{2s},}{}{A'C'_1\dots C'_q}{AC_1\dots C_q}
\csFparder{A'_1\cdots A'_{2s},}{}{C'_1\dots C'_q}{C_1\dots C_q} ) ,
\endEQs
where
$\csFder{}{A'_1\cdots A'_{2s},}{C_1\dots C_p}{C'_1\dots C'_p}$
denotes the complex conjugate of
$\sFder{}{A_1\cdots A_{2s},}{C'_1\dots C'_p}{C_1\dots C_p}$.
Due to the commutativity of partial derivatives,
$\sD{}{AA'}$ satisfies the identities
\EQ
\sD{A'}{(A} \sD{}{B)A'} =0, \qquad
\sD{A}{(A'} \sD{}{B')A}=0.
\label{sDids}
\endEQ

A locally constructed conserved current of
the massless spin $s$ field equations \eqref{Feq}
is a vector function $\dens{a}$
defined on some $J^q(\phi)$ satisfying
\EQ
\D{a}\dens{a}=0
\qquad\text{ on $R^{q}(\phi)$}.
\label{conscurr}
\endEQ
We refer to the integer $q$
as the order of $\dens{a}$.
The conserved current \eqref{conscurr} is trivial if
\EQ
\dens{a} =\D{b} \curl{ab}{}\quad
\text{on some $R^{p}(\phi)$},
\label{trivial}
\endEQ
where
$\curl{ab}{}=-\curl{ba}{}$
is some skew-tensor function on $J^p(\phi)$.
Two conserved currents are considered equivalent
if their difference is a trivial conserved current.
The smallest integer among the orders of all conserved currents
equivalent to $\dens{a}$ is called the order of
the equivalence class of conserved currents $\dens{a}$.
In spinor form,
a conserved current of order $q$ is a spinor function
$\dens{AA'} =\e{a}{AA'} \dens{a}$
satisfying
\EQ
\sD{}{AA'} \dens{AA'} =0
\qquad\text{ on $R^{q}(\phi)$},
\endEQ
which is trivial if
\EQ
\dens{AA'} = \sD{A}{B'}\curl{A'B'}{} +\sD{A'}{B} \tcurl{AB}{}\quad
\text{on some $R^{p}(\phi)$}
\label{scurl}
\endEQ
for some symmetric spinor functions $\curl{A'B'}{},\tcurl{AB}{}$
on $J^p(\phi)$
as determined from \eqref{trivial}
by the identity
$\curl{ab}{} \e{a}{AA'} \e{b}{BB'}
= \curl{A'B'}{}\vol{}{AB} + \tcurl{AB}{} \vol{}{A'B'}$.

\subsection{ Classification results }

Recall that in spinor form
a real conformal \Kvec/ \cite{Penrose,Wald}
$\othKV{a}{}(x)$ is represented by
a spinor function
$\othKV{AA'}{}(x)=\e{a}{AA'} \othKV{a}{}(x)$
satisfying
$\sder{(B'}{(B} \othKV{A')}{A)} =0$,
and a real conformal \KYten/ \cite{DietzRudiger}
$\KY{ab}{}(x)$ is represented by
a symmetric spinor function
$\KY{A'B'}{}(x)=\e{a A}{A'} \e{b}{AB'} \KY{ab}{}$
satisfying
$\sder{(C'}{C} \KY{A'B')}{}=0$,
with $\cKY{AB}{}(x)=\e{a A'}{A} \e{b}{BA'} \KY{ab}{}$.
Now define the conformally-weighted Lie derivative of
$\sF{}{A_1\cdots A_{2s}}(x)$ with respect to
a conformal \Kvec/ $\othKV{a}{}(x)$ by
\EQ
\sLie{\othKV{}{}} \sF{}{A_1\cdots A_{2s}}(x) =
( \Lie{\othKV{}{}} +\frac{1}{4}\div\othKV{}{} ) \sF{}{A_1\cdots A_{2s}}(x) ,
\label{sLie}
\endEQ
where $\Lie{\othKV{}{}}$ is the standard spinorial Lie derivative
(see \Ref{Penrose})
and $\div\othKV{}{} = \sder{}{EE'}\othKV{EE'}{}$.
A straightforward calculation shows that,
due to the linearity and conformal invariance of
the field equations \eqref{Feq},
$\sLie{\othKV{}{}} \sF{}{A_1\cdots A_{2s}}(x)$ is a solution
whenever $\sF{}{A_1\cdots A_{2s}}(x)$ is one.
Geometrically,
$\sLie{\othKV{}{}}\sF{}{A_1\cdots A_{2s}}(x)
= \Omega \Lie{\othKV{}{}}( \Omega^{-1}\sF{}{A_1\cdots A_{2s}}(x) )$,
where $\Omega(x)$ is the conformal factor of the spin metric
$\vol{AB}{} \rightarrow \Omega\vol{AB}{}$
under the one-parameter local conformal isometry of $M$
generated by $\othKV{}{}$
(see \Ref{Wald}),
with the corresponding conformal weight of $-1$
assigned to $\sF{}{A_1\cdots A_{2s}}$.
The Lie derivatives $\Lie{\othKV{}{}}$ and $\sLie{\othKV{}{}}$
have a natural lift to operators on $J^\infty(\phi)$
obtained by replacing $\der{AA'}$ with $\D{AA'}$.
Note that, consequently,
$\sLie{\othKV{}{}}$ is well defined on $R^\infty(\phi)$.

The elementary linear real-valued conserved currents
of the massless spin $s$ field equations \eqref{Feq}
are given by
\EQ
\curr{}{W}{}{AA'}(\phi,\bar\phi;\omega)
= \w{A'_1\cdots A'_{2s-1}}{A} \csF{}{A'A'_1\cdots A'_{2s-1}}
+ \cw{A_1\cdots A_{2s-1}}{A'} \sF{}{AA_1\cdots A_{2s-1}} ,
\label{Wcurr}
\endEQ
where $\w{A'_1\cdots A'_{2s-1}}{A}(x)$ is a spinor function
satisfying the adjoint spin $s$ field equations
\EQ
\sder{A(A'_1}{} \w{A'_2\cdots A'_{2s})}{A} =0 .
\label{adjointFeq}
\endEQ
It is straightforward to verify that \eqref{Wcurr}
is a conserved current of \eqref{Feq} due to \eqref{adjointFeq}.

The stress-energy currents, zilch currents, and chiral currents
in \Ref{maxwell} for electromagnetic fields
generalize to massless spin $s$ fields as follows:
\EQs
\curr{}{T}{}{AA'}(\phi,\bar\phi;\zeta)
= &&
2 \othKV{A_1A'_1}{} \cdots \othKV{A_{2s-1}A'_{2s-1}}{}
\sF{}{AA_1\cdots A_{2s-1}} \csF{}{A'A'_1\cdots A'_{2s-1}} ,
\label{Tcurr}\\
\curr{}{Z}{}{AA'}(\phi,\bar\phi;\zeta)
= &&
\i \othKV{A_1A'_1}{} \cdots \othKV{A_{2s-1}A'_{2s-1}}{}
( \csF{}{A'A'_1\cdots A'_{2s-1}}
\sLie{\othKV{}{}} \sF{}{AA_1\cdots A_{2s-1}}
\nonumber\\&&
- \sF{}{AA_1\cdots A_{2s-1}}
\sLie{\othKV{}{}} \csF{}{A'A'_1\cdots A'_{2s-1}} ) ,
\label{Zcurr}\\
\curr{}{V}{}{AA'}(\phi,\bar\phi;Y,\zeta)
= &&
\KY{(A'_1B'_1}{}\cdots \KY{A'_{2s}B'_{2s})}{}
\csFder{}{B'_1\cdots B'_{2s}}{}{A'_1A}
\sLie{\othKV{}{}} \csF{}{A'A'_2\cdots A'_{2s}}
\nonumber\\&&
+ \cKY{(A_1B_1}{} \cdots \cKY{A_{2s}B_{2s})}{}
\sFder{}{B_1\cdots B_{2s}}{}{A_1A'}
\sLie{\othKV{}{}} \sF{}{AA_2\cdots A_{2s}}
\nonumber\\&&
+\frac{2s+1}{4s+1} \Big(
\sder{}{AA'_1}( \KY{(A'_1B'_1}{}\cdots \KY{A'_{2s}B'_{2s})}{} )
\csF{}{B'_1\cdots B'_{2s}}
\sLie{\othKV{}{}} \csF{}{A'A'_2\cdots A'_{2s}}
\nonumber\\&&
+ \sder{}{A'A_1}( \cKY{(A_1B_1}{} \cdots \cKY{A_{2s}B_{2s})}{} )
\sF{}{B_1\cdots B_{2s}}{}
\sLie{\othKV{}{}} \sF{}{AA_2\cdots A_{2s}} \Big) .
\label{Vcurr}
\endEQs

\Proclaim{ Proposition 2.1. }{
For each integer $n\ge 0$,
and for nonzero real conformal \Kvec/s $\zeta$
and real conformal \KYten/s $Y$,
the spinor functions
\EQs
&& \curr{(n)}{T}{}{AA'}(\phi,\bar\phi;\zeta)
= \curr{}{T}{}{AA'}((\sLie{\zeta})^n\phi,(\sLie{\zeta})^n\bar\phi;\zeta) ,
\label{extTcurr}\\
&& \curr{(n)}{Z}{}{AA'}(\phi,\bar\phi;\zeta)
= \curr{}{Z}{}{AA'}((\sLie{\zeta})^n\phi,(\sLie{\zeta})^n\bar\phi;\zeta) ,
\label{extZcurr}\\
&& \curr{(n)}{V}{}{AA'}(\phi,\bar\phi;Y,\zeta)
= \curr{}{V}{}{AA'}((\sLie{\zeta})^n\phi,(\sLie{\zeta})^n\bar\phi;Y,\zeta)
\label{extVcurr}
\endEQs
yield non-trivial quadratic real-valued conserved currents of
the massless spin $s$ field equations \eqref{Feq}
of order $n$, $n+1$, $n+1$, respectively.
The currents \eqrefs{extTcurr}{extZcurr}
possess even parity under the duality symmetry \eqref{sduality},
while the currents \eqref{extVcurr}
possess odd parity. }
We refer to these as the (higher order)
stress-energy, zilch, and chiral currents, respectively.
A discussion of their physical nature is given in \secref{quantities}.

Higher order elementary currents
\EQ
\curr{(n)}{W}{}{AA'}(\phi,\bar\phi;\omega,\zeta)
= \curr{}{W}{}{AA'}((\sLie{\zeta})^n\phi,(\sLie{\zeta})^n\bar\phi;\omega) ,
\quad n>0, 
\label{extWcurr}
\endEQ
are equivalent to the currents
$\curr{}{W}{}{AA'}(\phi,\bar\phi;(-\sLie{\zeta})^n\omega)$,
which are of order $0$. 
Note here
$\sLie{\othKV{}{}} \w{A'_1\cdots A'_{2s-1}}{A}$
is a solution of \eqref{adjointFeq}
whenever $\w{A'_1\cdots A'_{2s-1}}{A}$ is one.

We now state the main classification results for conserved currents.

\Proclaim{ Theorem 2.2. }{
Every locally constructed conserved current \eqref{conscurr}
of order $q\ge 0$
of the massless spin $s\ge 1/2$ field equations \eqref{Feq}
is equivalent to a sum of
elementary currents \eqref{Wcurr},
stress-energy currents \eqref{extTcurr} with $0\le n\le q$,
zilch currents \eqref{extZcurr} with $0\le n\le q-1$,
and chiral currents \eqref{extVcurr} with $0\le n\le q-1$,
involving a solution $\omega$ of
the adjoint field equations \eqref{adjointFeq},
and conformal \Kvec/s $\zeta$ and conformal \KYten/s $Y$;
\EQs
\curr{}{}{}{AA'} \simeq
&&
\curr{}{W}{}{AA'}(\phi,\bar\phi;\omega)
+\sum_{n\le q} (
\sum_{\zeta} \pm \curr{(n)}{T}{}{AA'}(\phi,\bar\phi;\zeta)
+\sum_{\zeta} \pm \curr{(n-1)}{Z}{}{AA'}(\phi,\bar\phi;\zeta)
\nonumber\\&&
+\sum_{\zeta,Y} \pm \curr{(n-1)}{V}{}{AA'}(\phi,\bar\phi;Y,\zeta) ) .
\nonumber
\endEQs
}

\Proclaim{ Theorem 2.3. }{
Up to the addition of lower order currents,
the number of linearly independent equivalence classes of
stress-energy currents \eqref{extTcurr} of order $n$,
zilch currents \eqref{extZcurr} of order $n+1$,
and chiral currents \eqref{extVcurr} of order $n+1$
is given by
\EQs
&& \frac{1}{3} (s+n)^2 (2s+2n+1)^2 (4s+4n+1) ,\\
&& \frac{1}{3} (s+n+1)^2 (2s+2n+1)^2 (4s+4n+3) ,\\
&& \frac{2}{3} (n+1)(2n+3)(2s+n+1) (4s+2n+3) (4s+4n+5) .
\endEQs }

A construction for an explicit basis of
linearly independent conserved quantities
arising from these quadratic currents is outlined in \secref{quantities}.
We remark that in the case $s=1$ the count formulae in Theorem~2.3
reduce to the results obtained in \Refs{FushchichNikitin,maxwell}. 
(Note, in the former work,
the ``order of a current'' equals what we define as its weight;
see Corollary~4.7 in \secref{currents}.)

Underlying the classification of conserved currents in Theorem~2.2
is a corresponding set of conserved tensors
of a certain natural covariant form.
A locally constructed conserved tensor of
the massless spin $s$ field equations \eqref{Feq}
is a tensor function $\S{a_1\cdots a_r}{}$
defined on some $J^q(\phi)$ satisfying
\EQ
\D{a_1}\S{a_1\cdots a_r}{} =0
\qquad\text{ on $R^{q}(\phi)$}.
\label{constens}
\endEQ
Such a tensor $\S{a_1\cdots a_r}{}$
is trivial if it agrees on some $R^q(\phi)$
with a tensor that satisfies \eqref{constens}
identically on $J^{q+1}(\phi)$.
Two conserved tensors are considered equivalent
if their difference is a trivial conserved tensor.
The spinor form of a conserved tensor is given by
$\S{A'_1\cdots A'_r}{A_1 \cdots A_r}
= \e{a_1 A_1}{A'_1} \cdots \e{a_r A_r}{A'_r} \S{a_1\cdots a_r}{}$.

We obtain conserved tensors from
the quadratic conserved currents \eqsref{extTcurr}{extVcurr}
by first setting $\othKV{AA'}{}$ to be a constant spinorial vector
and $\KY{A'B'}{}$ to be a constant symmetric spinor,
then factoring out $\othKV{A}{A'},\KY{}{A'B'},\cKY{AB}{}$.
The stress-energy and zilch currents thereby yield
\EQs
\T{T}{A'_1\cdots A'_{2s}}{A_1\cdots A_{2s}}(\phi,\bar\phi)
= &&
2 \sF{}{A_1\cdots A_{2s}} \csF{A'_1\cdots A'_{2s}}{} ,
\label{Ttens}\\
\T{Z}{A'_1\cdots A'_{2s}B'}{A_1\cdots A_{2s}B}(\phi,\bar\phi)
= &&
\i \csF{(A'_1\cdots A'_{2s}}{} \sFder{}{A_1\cdots A_{2s}}{B')}{B}
-\i\sF{}{(A_1\cdots A_{2s}} \csFder{A'_1\cdots A'_{2s}}{}{B'}{B)} .
\label{Ztens}
\endEQs
Taking into account that $\KY{}{A'B'}$ is complex-valued,
after some algebraic manipulations
we find that the chiral current yields
\EQs
\T{V_+}{A'_1A'_2 B'_1C'_1\cdots B'_{2s}C'_{2s}}
{A_1A_2 B_1C_1\cdots B_{2s}C_{2s}}(\phi,\bar\phi)
= &&
\T{V}{A'_1A'_2 B'_1C'_1\cdots B'_{2s}C'_{2s}}
{A_1A_2 B_1C_1\cdots B_{2s}C_{2s}}(\bar\phi)
+\T{\bar V}{A'_1A'_2 B'_1C'_1\cdots B'_{2s}C'_{2s}}
{A_1A_2 B_1C_1\cdots B_{2s}C_{2s}}(\phi) ,
\label{V+tens}\\
\T{V_-}
{A'_1A'_2 B'_1C'_1\cdots B'_{2s}C'_{2s}}
{A_1A_2 B_1C_1\cdots B_{2s}C_{2s}}(\phi,\bar\phi)
= &&
\i\T{V}{A'_1A'_2 B'_1C'_1\cdots B'_{2s}C'_{2s}}
{A_1A_2 B_1C_1\cdots B_{2s}C_{2s}}(\bar\phi)
-\i\T{\bar V}{A'_1A'_2 B'_1C'_1\cdots B'_{2s}C'_{2s}}
{A_1A_2 B_1C_1\cdots B_{2s}C_{2s}}(\phi) ,
\label{V-tens}
\endEQs
where
\EQs
\T{V}{A'_1A'_2 B'_1C'_1\cdots B'_{2s}C'_{2s}}
{A_1A_2 B_1C_1\cdots B_{2s}C_{2s}}(\bar\phi)
= &&
\vol{B_1C_1}{}\cdots \vol{B_{2s}C_{2s}}{} (
\csFder{(B'_1\cdots B'_{2s}}{}{|A'_1|}{(A_1}
\csFder{C'_1\cdots C'_{2s})}{}{A'_2}{A_2)}
\nonumber\\&&\qquad
- \csFder{(B'_1\cdots B'_{2s}}{}{|A'_2|}{[A_2}
\csFder{C'_1\cdots C'_{2s})}{}{A'_1}{A_1]} ) .
\label{Vtens}
\endEQs
We refer to \eqsref{Ttens}{Vtens}
as the spin $s$ energy tensor, zilch tensor, and chiral tensors,
respectively.

\Proclaim{ Proposition 2.4. }{
The spinor functions \eqref{Ttens}, \eqref{Ztens}, \eqrefs{V+tens}{V-tens}
yield real-valued conserved tensors of
the massless spin $s$ field equations \eqref{Feq},
given by
\EQs
&&
\T{T}{a_1\cdots a_{2s}}{}(\phi,\bar\phi) ,\quad
\T{Z}{a_1\cdots a_{2s}b}{}(\phi,\bar\phi) ,
\label{TZtensors}\\
&&
\T{V_\pm}{a_1a_2 b_1c_1\cdots b_{2s}c_{2s}}{}(\phi,\bar\phi)
\label{Vtensors} .
\endEQs
On $R^\infty(\phi)$,
both the energy and zilch tensors \eqref{TZtensors}
are symmetric and trace-free on all their indices,
while the chiral tensors \eqref{Vtensors}
are symmetric on their first two indices
and skew on each of the last $2s$ pairs of indices,
symmetric under the interchange of all pairs of skew indices,
trace-free on the last $4s+1$ indices,
and satisfy the duality relation
\EQ
*\T{V_\pm}{a_1a_2 b_1c_1\cdots b_{2s}c_{2s}}{}(\phi,\bar\phi)
= \pm \T{V_\mp}{a_1a_2 b_1c_1\cdots b_{2s}c_{2s}}{}(\phi,\bar\phi)
\endEQ
where $*$ denotes the Hodge dual operator $\frac{1}{2}\vol{ab}{cd}$
defined in terms of the metric volume form
acting on a skew pair of indices. }

The conserved tensors \eqrefs{TZtensors}{Vtensors}
have higher order extensions
\EQs
&&
\T{T}{a_1\cdots a_{2s}}{}
((\sLie{\zeta})^n\phi,(\sLie{\zeta})^n\bar\phi) 
\label{extendedTtens}\\
&&
\T{Z}{a_1\cdots a_{2s}b}{}
((\sLie{\zeta})^n\phi,(\sLie{\zeta})^n\bar\phi) 
\label{extendedZtens}\\
&&
\T{V_\pm}{a_1a_2 b_1c_1\cdots b_{2s}c_{2s}}{}
((\sLie{\zeta})^n\phi,(\sLie{\zeta})^n\bar\phi) 
\label{extendedVtens}
\endEQs
for real conformal \Kvec/s $\zeta$,
for every $n\ge 0$. 
By taking $\zeta$ to be a constant \Kvec/
in \eqsref{extendedTtens}{extendedVtens}
and factoring it out,
one obtains higher rank energy tensors, zilch tensors, and chiral tensors
that are locally constructed
in a covariant fashion purely from the variables \eqref{symmdervar}
and the spin metric.
We will prove a classification result for such covariant conserved tensors
in \Ref{tensors}.

Finally, we remark that the extended energy tensor \eqref{Ttens}
also arises naturally from the so-called super-energy construction
\cite{Senovilla}
applied to the massless spin $s$ field equations \eqref{Feq}.
Further properties of \eqref{Ttens},
in particular a dominant energy condition,
can be obtained from the results in \Ref{Senovilla}.

\subsection{ Preliminaries }

Here we outline the main steps in the proofs of
the classification theorems of the previous section.
For any conserved current \eqref{conscurr} of order $q$,
it is convenient to consider an equivalent conserved current,
which we again denote by $\dens{AA'}$,
given by the standard integration by parts procedure \cite{Olver}
such that
\EQ
\D{AA'}\dens{AA'} =
\Q{A'_1\cdots A'_{2s-1}}{B} \csFder{}{B'A'_1\cdots A'_{2s-1},}{B'B}{}
+ \tQ{A_1\cdots A_{2s-1}}{B'} \sFder{}{BA_1\cdots A_{2s-1},}{BB'}{}
\label{charcurreq}
\endEQ
for some spinor functions
$\Q{A'_1\cdots A'_{2s-1}}{B},\tQ{A_1\cdots A_{2s-1}}{B'}$
on $J^{r}(\phi)$ for some $r\leq 2q$.
The pair
\EQ
\bigQ{}=(\Q{A'_1\cdots A'_{2s-1}}{B},\tQ{A_1\cdots A_{2s-1}}{B'})
\label{Q}
\endEQ
is referred to
as the {\it characteristic} of the conserved current $\dens{AA'}$
and the integer $r$ is called the order of $\bigQ$.
If $\H{AA'}{}$ is a conserved current equivalent to $\dens{AA'}$
then we call $\bigQ$ a characteristic {\it admitted} by $\H{AA'}{}$.
A conserved current $\dens{AA'}$ is equivalent to a real-valued current
only if it admits characteristic spinor functions that are complex
conjugates
$\tQ{A_1\cdots A_{2s-1}}{B'} = \cQ{A_1\cdots A_{2s-1}}{B'}$.
In this case we simply call $\Q{A'_1\cdots A'_{2s-1}}{B}$ 
the characteristic admitted by $\dens{AA'}$. 
Given a characteristic $\bigQ{}$,
one can employ a locally constructed integral formula
\cite{Olver,AncoBlumanII}
to obtain a conserved current $\dens{AA'}$
satisfying \eqref{charcurreq}
(namely, solving the divergence equations 
by an application of
the homotopy operator of the Euler-Lagrange complex \cite{Anderson}
on the jet space $J^\infty(\phi)$).

It can be straightforwardly shown that, for $s\ge 1$,
the gradient expressions
\EQ
\Q{A'_1\cdots A'_{2s-1}}{B} = \sD{(A'_1}{B} \XQ{A'_2\cdots A'_{2s-1})}
,\qquad
\tQ{A_1\cdots A_{2s-1}}{B'} = \sD{(A_1}{B'} \tXQ{A_2\cdots A_{2s-1})}
\label{gradientQ}
\endEQ
for any symmetric spinor functions
$\XQ{A'_1\cdots A'_{2s-2}}$, $\tXQ{A_1\cdots A_{2s-2}}$
on $J^{r-1}(\phi)$ yield a characteristic $\bigQ$
that determines a trivial conserved current \eqref{scurl}
with
$\curl{}{A'B'} = \XQ{A'_1\cdots A'_{2s-2}} \csF{}{A'B'A'_1\cdots A'_{2s-2}}$
and
$\tcurl{}{AB} = \tXQ{A_1\cdots A_{2s-2}} \sF{}{ABA_1\cdots A_{2s-2}}$
due to identities \eqref{sDids}.
Moreover, by the linearity of the field equations \eqref{Feq},
one can show that \cite{Olver}
a conserved current
with a characteristic $\bigQ$ that vanishes on some $R^q(\phi)$
is trivial.
Consequently,
we call a characteristic $\bigQ$ {\it trivial} if
it agrees with symmetrized spinor gradients \eqref{gradientQ}
when restricted to some $R^p(\phi)$.
Two characteristics are considered to be equivalent
if their difference is a trivial characteristic.
By the same proof as in the case $s=1$ given in \Ref{maxwell},
we now find the following relationship between
conserved currents and characteristics.

\Proclaim{Theorem 2.5. }{
There is a one-to-one correspondence between
equivalence classes of conserved currents
and equivalence classes of characteristics
for the massless spin $s\ge 1/2$ field equations \eqref{Feq}. }

Necessary conditions for a pair of spinor functions \eqref{Q} to satisfy
the characteristic equation \eqref{charcurreq}
can be obtained by the use of the standard Euler operator \cite{Olver}
that annihilates total divergence expressions on $J^\infty(\phi)$.
In particular,
one can show that the symmetrized divergences
$\sD{A(A'_1}{} \Q{A'_2\cdots A'_{2s})}{A}$
and $\sD{A'(A_1}{} \tQ{A_2\cdots A_{2s})}{A'}$
of a characteristic $\bigQ$
are proportional to the field equations \eqref{Feq}
and their derivatives,
and thus
all characteristics of order $r$ satisfy
\EQ
\sD{A(A'_1}{} \Q{A'_2\cdots A'_{2s})}{A} =0 ,\qquad
\sD{A'(A_1}{} \tQ{A_2\cdots A_{2s})}{A'} =0
\qquad\eqtext{on $R^{r}(\phi)$} .
\label{Qadsymmeq}
\endEQ
These equations are the adjoint of the determining equations
\EQ
\sD{A'_1}{A} \X{A'_1\cdots A'_{2s}} =0 ,\qquad
\sD{A_1}{A'} \tX{A_1\cdots A_{2s}} =0
\qquad\eqtext{on $R^{r}(\phi)$}
\label{Xsymmeq}
\endEQ
for symmetries
$\bigX=
\X{A'_1\cdots A'_{2s}} \csFparder{}{A'_1\cdots A'_{2s}}{}{}
+ \tX{A_1\cdots A_{2s}} \sFparder{}{A_1\cdots A_{2s}}{}{}$
of massless spin $s\ge 1/2$ fields \cite{maxwellsymm}.
We refer to \eqref{Qadsymmeq} as the {\it adjoint symmetry equations}
and we call functions
\EQ
\bigP=(\P{A'_1\cdots A'_{2s-1}}{B},\tP{A_1\cdots A_{2s-1}}{B'})
\label{P}
\endEQ
defined on $J^r(\phi)$
an {\it adjoint symmetry} of order $r$
of massless spin $s\ge 1/2$ fields
if \eqref{P} satisfies
\EQ
\sD{B(A'_1}{} \P{A'_2\cdots A'_{2s})}{B} =0 ,\qquad
\sD{B'(A_1}{} \tP{A_2\cdots A_{2s})}{B'} =0
\qquad\eqtext{on $R^{r}(\phi)$}.
\label{Pdeteq}
\endEQ
(Note that, under complex conjugation,
solutions of the first equation go into solutions of the second equation,
and conversely.)
It follows from equation \eqref{gradientQ} that
if $s\ge 1$ the symmetrized spinor gradients
\EQ
\P{A'_1\cdots A'_{2s-1}}{B} = \sD{(A'_1}{B} \XQ{A'_2\cdots A'_{2s-1})}
,\qquad
\tP{A_1\cdots A_{2s-1}}{B'} = \sD{(A_1}{B'} \tXQ{A_2\cdots A_{2s-1})}
\label{gaugeP}
\endEQ
for any symmetric spinor functions
$\XQ{A'_1\cdots A'_{2s-2}}$, $\tXQ{A_1\cdots A_{2s-2}}$
on $J^{r-1}(\phi)$
trivially satisfy \eqref{Pdeteq}.
We call $\bigP$ an {\it adjoint gauge symmetry}
if it agrees with \eqref{gaugeP} on $R^{r-1}(\phi)$,
and we consider two adjoint symmetries to be equivalent
if their difference is an adjoint gauge symmetry.
The order $r$ of an adjoint symmetry $\bigP$ is called {\it minimal}
if it is the smallest among the orders of all adjoint symmetries
equivalent to $\bigP$.
If $\bigP$ is not equivalent to an adjoint gauge symmetry
then we call $\bigP$ non-trivial.
Note that equations \eqref{Pdeteq} do not admit
any adjoint gauge symmetry solutions when $s=1/2$.

Thus, all characteristics $\bigQ$ of conserved currents
of the massless spin $s\ge 1/2$ field equations \eqref{Feq}
are adjoint symmetries.
However, an adjoint symmetry $\bigP$ is not a characteristic
of a conserved current
unless it satisfies the characteristic equation \eqref{charcurreq}.
This requires certain differential conditions 
\cite{Olver,AncoBluman1,AncoBlumanII}
to hold on $\bigP$ in addition to \eqref{Pdeteq}.
Indeed, massless spin $s\ge 1/2$ fields
admit non-trivial adjoint symmetries
that fail to be equivalent to characteristics,
similarly to the case $s=1$ treated in \Ref{maxwell}.
Nevertheless, due to the linearity of the field equations \eqref{Feq},
this difficulty can be by-passed
by employing a variant of the standard integral formula
for constructing a conserved current \eqref{charcurreq}
from its characteristic $\bigQ$.

Let
\EQ
\H{}{AA'}(\bigP) =
\int_0^1 d\lambda \big(
\csF{}{A'A'_1\cdots A'_{2s-1}}
\P{A'_1\cdots A'_{2s-1}}{A}(x,\lambda\phi,\lambda\bar\phi)
+ \sF{}{AA_1\cdots A_{2s-1}}
\tP{A_1\cdots A_{2s-1}}{A'}(x,\lambda\phi,\lambda\bar\phi)
\big) ,
\label{currformula}
\endEQ
where
$\bigP=( \P{A'_1\cdots A'_{2s-1}}{A}, \tP{A_1\cdots A_{2s-1}}{A'} )$
is a pair of spinor functions defined on some $J^q(\phi)$,
with $\phi,\bar\phi$
standing collectively for all the variables
$\sFder{}{A_1\cdots A_{2s},}{C'_1\cdots C'_p}{C_1\cdots C_p}$,
$\csFder{A'_1\cdots A'_{2s},}{}{C'_1\cdots C'_p}{C_1\cdots C_p}$,
$p\ge 0$.

\Proclaim{ Proposition 2.6. }{
Let $\bigP= (\P{A'_1\cdots A'_{2s-1}}{A}, \tP{A_1\cdots A_{2s-1}}{A'} )$
be an adjoint symmetry of order $r$ of
the massless spin $s\ge 1/2$ field equations \eqref{Feq}.
Then $\H{AA'}{}(\bigP)$ is a conserved current of order $r$.
If $\bigP$ is equivalent to the characteristic $\bigQ$
of a conserved current $\dens{AA'}$,
then the current $\H{AA'}{}(\bigP)$ is equivalent to $\dens{AA'}$.
In particular, if $\bigP$ is equivalent to a trivial characteristic,
then $\H{AA'}{}(\bigP)$ is a trivial current. }

\Proclaim{Corollary 2.7. }{
If an adjoint symmetry $\bigP$ is not equivalent
to a characteristic $\bigQ$ admitted by
the conserved current $\H{AA'}{}(\bigP)$,
then $\bigP$ is not equivalent to the characteristic of
any conserved current. }

The proof of these results is analogous to that presented in \Ref{maxwell}
for the case $s=1$.
We remark that the conserved current formula \eqref{currformula} originates
\cite{AncoBluman1,AncoBluman2}
from the conservation law identity
\EQ
\coD{BB'}(
\P{A'_2\cdots A'_{2s}}{B} \X{B'A'_2\cdots A'_{2s}}
+ \tP{A_2\cdots A_{2s}}{B'} \tX{BA_2\cdots A_{2s}} )=0
\qquad\text{on $R^\infty(\phi)$},
\endEQ
which holds due to the adjoint relation between
equations \eqrefs{Qadsymmeq}{Xsymmeq}.
In particular,
we obtain \eqref{currformula}
by first letting
$\X{B'A'_2\cdots A'_{2s}}=\csF{}{B'A'_2\cdots A'_{2s}}$
and
$\tX{BA_2\cdots A_{2s}} = \sF{}{BA_2\cdots A_{2s}}$
be the scaling symmetry of the field equations \eqref{Feq},
then replacing the variables $\phi,\bar\phi$
by the one-parameter scaling $\lambda\phi,\lambda\bar\phi$,
and finally dividing by $\lambda$
and integrating from $\lambda=0$ to $\lambda=1$.
In comparison,
the standard integral formula arises in a similar manner
\cite{AncoBlumanII}
from a well-known identity \cite{Olver,Anderson}
relating the Euler operator and linearization operator
on $J^\infty(\phi)$
(namely, the first variational formula of the calculus of variations),
applied to the characteristic equation \eqref{charcurreq}.

As a consequence of Theorem~2.5 and Proposition~2.6,
we are able to establish the classification result of Theorem~2.2
for all conserved currents of the massless spin $s\ge 1/2$ field equations
by the following steps:
\vskip0pt
(i) classify up to equivalence all adjoint symmetries of
massless spin $s\ge 1/2$ fields;
\hfill\vskip0pt
(ii) use integral formula \eqref{currformula} to construct
the conserved currents arising from
the equivalence classes of adjoint symmetries found in step (i);
\hfill\vskip0pt
(iii) calculate a characteristic for each equivalence class of
conserved currents found in step (ii);
\hfill\vskip0pt
(iv) classify the equivalences classes of characteristics
found in step (iii).

\section{Classification of adjoint symmetries}
\label{adjointsymms}

We solve the adjoint symmetry equations
\EQ
\sD{A(A'_1}{} \P{A'_2\cdots A'_{2s})}{A} =0
\qquad\eqtext{on $R^{r}(\phi)$}
\label{adjsymmeq}
\endEQ
by spinorial methods.
To begin, we summarize some useful preliminary results.

In flat spacetime,
a \Kspin/ $\KS{A_1 \cdots A_k A'_1 \cdots A'_l}{}(x)$ of type $(k,l)$
is a solution of the conformally invariant equation
\cite{Penrose}
\EQ
\sder{(B'|(B}{} \KS{A_1 \cdots A_k) |A'_1 \cdots A'_l)}{} =0 .
\label{KSeq}
\endEQ
Complex conjugation of a type $(k,l)$ \Kspin/
yields a type $(l,k)$ \Kspin/.
Note that a type $(1,1)$ \Kspin/ corresponds to
a complex-valued conformal \Kvec/ $\othKV{a}{} = \inve{a}{AA'} \KS{AA'}{}$,
and a type $(0,2)$ \Kspin/ corresponds to a
conformal \KYten/ $\KY{ab}{} = \inve{a}{AA'} \inve{b A}{B'}\KS{A'B'}{}$
which is self-dual,
$*\KY{ab}{}=i\KY{ab}{}$,
where $*$ denotes the Hodge dual operator.
We refer to a type $(k,k)$ \Kspin/ as a rank $k$ conformal \Kten/,
and a type $(0,2l)$ \Kspin/ as a rank $l$ self-dual conformal \KYten/.
A rank $k$ conformal \Kten/ that is real,
\ie/, invariant under complex conjugation,
will be called a type $(k,k)_R$ \Kspin/.
\Kspin/s in flat spacetime can be factored
into sums of symmetrized products of
type $(0,1)$ and $(1,0)$ \Kspin/s,
which corresponds to the factorization of
principal parts of trace-free symmetric twistors
in terms of elementary twistors \cite{Penrose}.
The following Lemma,
which is a consequence of this factorization property,
will be pivotal in our analysis of adjoint symmetries
and conserved currents.

\Proclaim{Lemma~3.1. }{
A symmetric spinor function
$\KS{A_1 \cdots A_k A'_1 \cdots A'_k}{}(x)$
is a Killing spinor of type $(k,k)$
if and only if it can be expressed as
a sum of symmetrized products of conformal \Kvec/s
$\KS{A'_1 \cdots A'_k}{A_1 \cdots A_k}
= \sum_\zeta \othKV{A'_1}{(A_1} \cdots \othKV{A'_k}{A_k)}$.
A symmetric spinor function
$\KS{A'_1 \cdots A'_{2l}}{}(x)$
is a Killing spinor of type $(0,2l)$
if and only if it can be expressed as
a sum of symmetrized products of self-dual conformal \KYten/s
$\KS{A'_1 \cdots A'_{2l}}{}
= \sum_Y \KY{(A'_1A'_2}{} \cdots \KY{A'_{2l-1}A'_{2l})}{}$.
More generally,
a symmetric spinor function
$\KS{A_1 \cdots A_{k} A'_1 \cdots A'_{k+2l}}{}(x)$
is a Killing spinor of type $(k,k+2l)$
if and only if it can be expressed as
a sum of symmetrized products of
rank $k$ conformal \Kten/s
and rank $l$ self-dual conformal \KYten/s.
There are
\EQs
(k+1) (k+2) (k+2l+1) (k+2l+2) (2k+2l+3)/12
\nonumber
\endEQs
linearly independent \Kspin/s of type $(k,k+2l)$
over the complex numbers.}

It is straightforward to verify that,
due to the invariance of the \Kspin/ equation \eqref{KSeq}
under local conformal scalings of the Minkowski metric,
the Lie derivative
$\Lie{\zeta} \KS{A_1 \cdots A_k A'_1 \cdots A'_l}{}$
of a \Kspin/
$\KS{A_1 \cdots A_k A'_1 \cdots A'_l}{}$
with respect to a conformal \Kvec/ $\zeta$
is again a \Kspin/ of the same type.
Recall that,
if $\sF{}{A_1\cdots A_{2s}}(x)$ is a solution of
the massless spin $s$ field equations \eqref{Feq},
then so is its conformally-weighted Lie derivative
$\sLie{\othKV{}{}} \sF{}{A_1\cdots A_{2s}}(x)$
with respect to a conformal \Kvec/ $\othKV{}{}$.
For convenience we write
\EQ\label{nfoldF}
\snF{n}{\zeta}{}{A_1\cdots A_{2s}}
= (\sLie{\zeta})^n \sF{}{A_1\cdots A_{2s}} ,
\qquad
\csnF{n}{\zeta}{}{A'_1\cdots A'_{2s}}
= (\sLie{\zeta})^n \csF{}{A'_1\cdots A'_{2s}} ,
\qquad n\ge 0 .
\endEQ

A spinor function $\w{A'_1\cdots A'_{2s-1}}{A}(x)$
satisfying the adjoint spin $s$ field equations \eqref{adjointFeq}
is called an {\it elementary} adjoint symmetry,
which we denote by
$\W{A'_1\cdots A'_{2s-1}}{A}(\omega) = \w{A'_1\cdots A'_{2s-1}}{A}$.
To proceed,
we present the spin $s\ge 1/2$ analogs of
the non-elementary adjoint symmetries found in \Ref{maxwell}
for Maxwell's equations.

\Proclaim{Proposition~3.2 } {
Let
$\othKV{AA'}{}$,
$\KV{A_1\cdots A_{2s-1} A'_1\cdots A'_{2s-1}}{}$,
$\othKY{A'_1\cdots A'_{4s}}{}$
be Killing spinors.
Then, for each integer $n\ge 0$,
the spinor functions
\EQs
\zeroP{A'_1\cdots A'_{2s-1}}{A}(\phi;n,\zeta,\xi)
= &&
\KV{A_1\cdots A_{2s-1} A'_1\cdots A'_{2s-1}}{}
\snF{n}{\zeta}{}{A A_1\cdots A_{2s-1}},
\label{Uadjsymm}\\
\firstP{A'_1\cdots A'_{2s-1}}{A}(\bar\phi;n,\zeta,\Upsilon)
= &&
\othKY{A'A'_1\cdots A'_{2s-1}B'_1\cdots B'_{2s}}{}
\D{AA'}\csnF{n}{\zeta}{}{B'_1\cdots B'_{2s}}
\nonumber\\&&\qquad
+\frac{2s+1}{4s+1}
\sder{}{AA'} \othKY{A'A'_1\cdots A'_{2s-1}B'_1\cdots B'_{2s}}{}
\csnF{n}{\zeta}{}{B'_1\cdots B'_{2s}}
\label{Vadjsymm}
\endEQs
are adjoint symmetries of order $n$ and $n+1$, respectively.
Furthermore,
these are equivalent to non-trivial adjoint symmetries
whose respective highest order terms are given by
\EQs
&&
(-1)^n \KV{A_1\cdots A_{2s-1} (A'_1\cdots A'_{2s-1}}{}
\othKV{C_1'}{C_1}\cdots\othKV{C'_n)}{C_n}
\sFder{}{A A_1\cdots A_{2s-1}}{C_1\cdots C_n}{C'_1\cdots C'_n} , \quad
\label{highestorderU}
\\&&
(-1)^n \othKY{(A'A'_1\cdots A'_{2s-1}B'_1\cdots B'_{2s}}{}
\othKV{C_1'}{C_1}\cdots\othKV{C'_n)}{C_n} \vol{BA}{}
\csFder{}{B'_1\cdots B'_{2s}}{BC_1\cdots C_n}{A'C'_1\cdots C'_n} ,
\label{highestorderV}
\endEQs
provided that $\zeta,\xi,\Upsilon$ are non-zero. }

We next state the classification result for adjoint symmetries
of all massless spin $s\ge 1/2$ fields.

\Proclaim{Theorem 3.3. }{
Every adjoint symmetry $\P{A'_1\cdots A'_{2s-1}}{A}$ of order $r$
is equivalent to a sum of
an elementary adjoint symmetry $\W{A'_1\cdots A'_{2s-1}}{A}(\omega)$
and linear adjoint symmetries
$\zeroP{A'_1\cdots A'_{2s-1}}{A}(\phi;n,\zeta,\xi)$,
$0\leq n\leq r$,
and
$\firstP{A'_1\cdots A'_{2s-1}}{A}(\bar\phi;n,\zeta,\Upsilon)$,
$0\leq n\leq r-1$,
involving
rank $2s-1$ conformal \Kten/s $\xi$,
rank $2s$ self-dual conformal \KYten/s $\Upsilon$,
and conformal \Kvec/s $\zeta$;
\EQs
\P{A'_1\cdots A'_{2s-1}}{A} \simeq
&&
\W{A'_1\cdots A'_{2s-1}}{A}(\omega)
+\sum_{n\le r} (
\sum_{\zeta,\xi}
\zeroP{A'_1\cdots A'_{2s-1}}{A}(\phi;n,\zeta,\xi)
\nonumber\\&&\qquad
+ \sum_{\zeta,\Upsilon}
\firstP{A'_1\cdots A'_{2s-1}}{A}(\bar\phi;n-1,\zeta,\Upsilon) ) .
\endEQs
}

The proof of this theorem involves four main steps:
First, one solves for the highest order terms
in $\P{A'_1\cdots A'_{2s-1}}{A}$
by means of a linearization technique.
Next, using Proposition~3.2 combined with Lemma~3.1,
one shows that these terms in $\P{A'_1\cdots A'_{2s-1}}{A}$
agree with the highest order terms
\eqref{highestorderU} and \eqref{highestorderV}
in the linear adjoint symmetries
\eqrefs{Uadjsymm}{Vadjsymm}
up to an adjoint gauge symmetry.
Hence, after subtracting these adjoint symmetries
from $\P{A'_1\cdots A'_{2s-1}}{A}$,
one obtains a solution of lower order.
Finally, one completes the proof by applying
a descent argument with respect to the order of
the adjoint symmetry.
The details of these steps are similar to those in
the case $s=1$ in \Ref{maxwell}
and therefore will not be repeated here.
We remark that the first step relies on
the local solvability \cite{Olver}
of the field equations \eqref{Feq},
\ie/, for each point \eqref{Rpoint} in $R^q(\phi)$
there is a solution
$\sFsol{}{A_1\cdots A_{2s}}(x)$ of \eqref{Feq}
such that
$\sFsol{}{A_1\cdots A_{2s}}(x_0)= \sF{}{A_1\cdots A_{2s}}$
and
$\sder{(C'_1}{(C_1} \cdots \sder{C'_p)}{C_p}
\sFsol{}{A_1\cdots A_{2s})}(x_0)
= \sFder{}{A_1\cdots A_{2s}}{C'_1\cdots C'_p}{C_1\cdots C_p}$,
$1\le p\leq q+1$.

\section{ Classification of conserved currents }
\label{currents}

We now proceed to determine the real-valued conserved currents
arising from the classification of adjoint symmetries
in Theorem~3.3.
To begin,
a conserved current $\dens{AA'}$ of order $q$ is called linear/quadratic
if it can be expressed as a homogeneous linear/quadratic polynomial
in the variables
$\sFder{}{A_1\cdots A_{2s},}{C_1'\cdots C_p'}{C_1\cdots C_p}$, $0\le p\le
q$,
and their complex conjugates.
Let the {\it weight} of a monomial
be the sum of the orders of these variables,
and let the weight of a linear/quadratic current $\dens{AA'}$
be the maximum of the weights of all monomials in $\dens{AA'}$.
This weight is called {\it minimal} if
it is the smallest among the weights of
all quadratic currents equivalent to $\dens{AA'}$.
By the same proof as in the case $s=1$ in \Ref{maxwell},
we obtain the following result.

\Proclaim{ Proposition~4.1. }{
A conserved current $\dens{AA'}$ is equivalent to
a linear/quadratic current of minimal weight $w$
if and only if $\dens{AA'}$ admits a characteristic
that is equivalent to an elementary/linear adjoint symmetry
$\P{A'_1\cdots A'_{2s-1}}{A}$
of minimal order $w$. }

For an elementary adjoint symmetry
$\P{A'_1\cdots A'_{2s-1}}{A} =
\W{A'_1\cdots A'_{2s-1}}{A}(\omega)$,
the real-valued conserved current arising from Proposition~2.6
is given by
\EQ
\WH{}{AA'}[\omega] =
\csF{}{A'A'_1\cdots A'_{2s-1}}{} \w{A'_1\cdots A'_{2s-1}}{A}
+ c.c. ,
\label{Wdensity}
\endEQ
where $c.c.$ denotes the complex conjugate term. 
Let
$\othKV{CC'}{}$
be a real conformal \Kvec/,
$\KV{A_1\cdots A_{2s-1} A'_1\cdots A'_{2s-1}}{}$
be a rank $2s-1$ real conformal \Kten/,
$\othKY{A'_1\cdots A'_{4s}}{}$
be a rank $2s$ self-dual conformal \KYten/.
For linear adjoint symmetries
$\P{A'_1\cdots A'_{2s-1}}{A} =
\zeroP{A'_1\cdots A'_{2s-1}}{A}(\phi;n,\zeta,\xi)$,
$\P{A'_1\cdots A'_{2s-1}}{A} =
\i \zeroP{A'_1\cdots A'_{2s-1}}{A}(\phi;n,\zeta,\xi)$
of order $n$,
and
$\P{A'_1\cdots A'_{2s-1}}{A} =
\firstP{A'_1\cdots A'_{2s-1}}{A}(\bar\phi;n,\zeta,\Upsilon)$
of order $n+1$, $n\ge 0$,
the real-valued conserved currents arising from Proposition~2.6
are given by
\EQs
\TH{n}{}{AA'}[\xi,\zeta]
&=&
\frac{1}{2}
\csF{}{A'A'_1\cdots A'_{2s-1}}
\zeroP{A'_1\cdots A'_{2s-1}}{A}(\phi;n,\zeta,\xi)
+ c.c. ,
\label{Tdensity}\\
\ZH{n}{}{AA'}[\xi,\zeta]
&=&
\frac{\i}{2}
\csF{}{A'A'_1\cdots A'_{2s-1}}
\zeroP{A'_1\cdots A'_{2s-1}}{A}(\phi;n,\zeta,\xi)
+ c.c. ,
\label{Zdensity}\\
\VH{n}{}{AA'}[\Upsilon,\zeta]
&=&
\frac{1}{2}
\csF{}{A'A'_1\cdots A'_{2s-1}}
\firstP{A'_1\cdots A'_{2s-1}}{A}(\bar\phi;n,\zeta,\Upsilon)
+c.c. .
\label{Vdensity}
\endEQs


\Proclaim{ Lemma~4.2. } {
The conserved currents
\eqref{Wdensity}, \eqref{Tdensity}, \eqref{Zdensity}, \eqref{Vdensity}
admit the respective characteristics
\EQs
\WQ{A'_1\cdots A'_{2s-1}}{A}[\omega]
&=&
\w{A'_1\cdots A'_{2s-1}}{A} ,
\label{Wchar}\\
\TQ{n}{A'_1\cdots A'_{2s-1}}{A}[\xi,\zeta]
&=&
\frac{1+(-1)^n}{2}
\zeroP{A'_1\cdots A'_{2s-1}}{A}(\phi;n,\zeta,\xi)
\nonumber\\&&\qquad
+\frac{(-1)^n}{2} \sum_{p=1}^n \C{n}{p}
\zeroP{A'_1\cdots A'_{2s-1}}{A}(\phi;n-p,\zeta,(\Lie{\zeta})^p\xi) ,
\label{Tchar}\\
\ZQ{n}{A'_1\cdots A'_{2s-1}}{A}[\xi,\zeta]
&=&
\i \frac{1-(-1)^n}{2}
\zeroP{A'_1\cdots A'_{2s-1}}{A}(\phi;n,\zeta,\xi)
\nonumber\\&&\qquad
-\i\frac{(-1)^n}{2} \sum_{p=1}^n \C{n}{p}
\zeroP{A'_1\cdots A'_{2s-1}}{A}(\phi;n-p,\zeta,(\Lie{\zeta})^p\xi) ,
\label{Zchar}\\
\VQ{n}{A'_1\cdots A'_{2s-1}}{A}[\Upsilon,\zeta]
&=&
\frac{1-(-1)^n}{2}
\firstP{A'_1\cdots A'_{2s-1}}{A}(\bar\phi;n,\zeta,\Upsilon)
\nonumber\\&&\qquad
-\frac{(-1)^n}{2} \sum_{p=1}^n \C{n}{p}
\firstP{A'_1\cdots A'_{2s-1}}{A}(\bar\phi;n-p,\zeta,(\Lie{\zeta})^p\Upsilon)
.
\label{Vchar}
\endEQs
These characteristics are adjoint symmetries of order
$q_W,q_T,q_Z,q_V$, where
\EQs
&& q_T=n,  \eqtext{ if $n$ is even, } \quad
q_T <n,  \eqtext{ if $n$ is odd, }
\label{TQorder}\\
&& q_Z=n,  \eqtext{ if $n$ is odd, }\quad
q_Z <n,  \eqtext{ if $n$ is even, }
\label{ZQorder}\\
&& q_V=n+1,  \eqtext{ if $n$ is odd, }\quad
q_V <n+1,  \eqtext{ if $n$ is even, }
\label{VQorder}
\endEQs
and $q_W=0$.
In particular,
\eqref{Tchar} for $n=2r=q_T$,
\eqref{Zchar} for $n=2r+1=q_Z$,
\eqref{Vchar} for $n=2r+2=q_V$, $r\ge 0$,
are respectively equivalent to adjoint symmetries
with the highest order terms
\EQs
&&
\KV{A_1\cdots A_{2s-1} (A'_1\cdots A'_{2s-1}}{}
\othKV{C_1'}{C_1}\cdots\othKV{C'_{2r})}{C_{2r}}
\sFder{}{A A_1\cdots A_{2s-1}}{C_1\cdots C_{2r}}{C'_1\cdots C'_{2r}} ,
\label{highestorderTchar}\\
&&
-\i \KV{A_1\cdots A_{2s-1} (A'_1\cdots A'_{2s-1}}{}
\othKV{C_1'}{C_1}\cdots\othKV{C'_{2r+1})}{C_{2r+1}}
\sFder{}{A A_1\cdots A_{2s-1}}{C_1\cdots C_{2r+1}}{C'_1\cdots C'_{2r+1}} ,
\label{highestorderZchar}\\
&&
-\othKY{(A'A'_1\cdots A'_{2s-1}B'_1\cdots B'_{2s}}{}
\othKV{C_1'}{C_1}\cdots\othKV{C'_{2r+1})}{C_{2r+1}} \vol{BA}{}
\csFder{}{B'_1\cdots B'_{2s}}{BC_1\cdots C_{2r+1}}{A'C'_1\cdots C'_{2r+1}} ,
\label{highestorderVchar}
\endEQs
provided $\zeta,\xi,\Upsilon$ are non-zero.
}

The proof of Lemma~4.2 proceeds along the same steps
as in the case $s=1$ in \Ref{maxwell},
which we outline here in brief.
First, write
\EQ
\FE{A'}{A_1\cdots A_{2s-1}}(\phi)
= \sFder{}{A_1\cdots A_{2s-1} A,}{AA'}{} .
\label{FEvar}
\endEQ
This spinor function satisfies the Lie derivative identity
\EQ
\FE{A'}{A_1\cdots A_{2s-1}}(\sLie{\zeta}\phi)
=( \sLie{\zeta} +\frac{1}{2}\div\zeta )\FE{A'}{A_1\cdots A_{2s-1}}(\phi) .
\label{LieFE}
\endEQ
Similar identities hold for
the linear adjoint symmetries \eqrefs{Uadjsymm}{Vadjsymm},
\EQs
&&
\zeroP{A'_1\cdots A'_{2s-1}}{A}(\sLie{\zeta}\phi;n,\zeta,\xi)
=
\sLie{\zeta} \zeroP{A'_1\cdots A'_{2s-1}}{A}(\phi;n,\zeta,\xi)
- \zeroP{A'_1\cdots A'_{2s-1}}{A}(\phi;n,\zeta,\Lie{\zeta}\xi) ,
\label{zeroPLie}\\
&&
\firstP{A'_1\cdots A'_{2s-1}}{A}(\sLie{\zeta}\bar\phi;n,\zeta,\Upsilon)
=
\sLie{\zeta} \firstP{A'_1\cdots A'_{2s-1}}{A}(\bar\phi;n,\zeta,\Upsilon)
- \firstP{A'_1\cdots A'_{2s-1}}{A}(\bar\phi;n,\zeta,\Lie{\zeta}\Upsilon) ,
\label{firstPLie}
\endEQs
where, recall,
the Lie derivatives of the \Kspin/s $\xi$ and $\Upsilon$
are again \Kspin/s.
To proceed with the computation of the characteristics
\eqsref{Wchar}{Vchar},
we apply $\D{AA'}$ to obtain the total divergence of
the conserved currents \eqsref{Tdensity}{Vdensity}
and use the equations
\EQs
\sD{A(A'_1}{}\zeroP{A'_2\cdots A'_{2s})}{A}(\phi;n,\zeta,\xi)
= &&
\KV{A_1\cdots A_{2s-1} (A'_1\cdots A'_{2s-1}}{}
\FE{A'_{2s})}{A_1\cdots A_{2s-1}}(\snF{n}{\zeta}{}{}) ,
\\
\sD{A(A'_1}{}\firstP{A'_2\cdots A'_{2s})}{A}(\bar\phi;n,\zeta,\Upsilon)
= &&
\othKY{A'_1B'_1\cdots A'_{2s}B'_{2s}}{}
\sD{}{AB'_1} \cFE{A}{B'_2\cdots B'_{2s}}(\csnF{n}{\zeta}{}{})
\nonumber\\&&
+\frac{2s}{4s+1} \sder{}{AB'_1} \othKY{A'_1B'_1\cdots A'_{2s}B'_{2s}}{}
\cFE{A}{B'_2\cdots B'_{2s}}(\csnF{n}{\zeta}{}{}) .
\endEQs
Next we substitute identity \eqref{LieFE}
and integrate by parts to move all Lie derivatives from
$\FE{A'}{A_2\cdots A_{2s}}(\phi)$
and $\cFE{A}{B'_2\cdots B'_{2s}}{}(\bar\phi)$
onto the adjoint symmetries
$\zeroP{A'_2\cdots A'_{2s}}{A}(\phi;n,\zeta,\xi)$
and $\firstP{A'_2\cdots A'_{2s}}{A}(\bar\phi;n,\zeta,\Upsilon)$.
Then we repeatedly use the Lie derivative identities
\eqrefs{zeroPLie}{firstPLie},
which leads to the characteristic equation \eqref{charcurreq}
for the currents \eqsref{Tdensity}{Vdensity}
where $\bigQ$ is given by \eqsref{Tchar}{Vchar}.
Finally,
the characteristic \eqref{Wchar} is obtained immediately
from the total divergence of the current \eqref{Wdensity}.

From Lemma~4.2 combined with Proposition~4.1 and Theorem~2.5,
we now have the following classification result
for conserved currents.

\Proclaim{Theorem 4.3. }{
Every conserved current of
the massless spin $s\ge 1/2$ field equations \eqref{Feq}
is equivalent to
the sum of a linear current and a quadratic current.
The equivalence classes of linear currents
are represented by the currents
\EQ
\WH{AA'}{}[\omega] ,
\label{Wspan}
\endEQ
where $\omega$ satisfies the adjoint spin $s$ field equations
\eqref{adjointFeq}.
The equivalence classes of
quadratic currents of weight at most $w$
are represented by sums of the currents
\EQs
&& \TH{2r}{AA'}{}[\xi,\zeta],
\quad 0\leq r\leq [w/2],
\label{Tspan}\\
&& \ZH{2r+1}{AA'}{}[\xi,\zeta],
\quad 0\le r\leq [(w-1)/2],
\label{Zspan}\\
&& \VH{2r+1}{AA'}{}[\Upsilon,\zeta],
\quad 0\le r\leq [w/2]-1 ,
\label{Vspan}
\endEQs
involving
real conformal \Kvec/s $\zeta$,
rank $2s-1$ real conformal \Kten/s $\xi$,
and rank $2s$ self-dual conformal \KYten/s $\Upsilon$
for each $r$.
In particular,
up to quadratic currents of lower weight,
a quadratic current of minimal even weight $w=2r$ is equivalent to
a sum of currents
\EQs
\dens{AA'} \simeq
\sum_{\zeta,\xi} \pm \TH{w}{AA'}{}[\xi,\zeta]
+\sum_{\zeta,\Upsilon} \pm \VH{w-1}{AA'}{}[\Upsilon,\zeta] ,
\nonumber
\endEQs
and a quadratic current of minimal odd weight $w=2r+1$ is equivalent to
a sum of currents
\EQs
\dens{AA'} \simeq
\sum_{\zeta,\xi}
\pm \ZH{w}{AA'}{}[\xi,\zeta] .
\nonumber
\endEQs
 }

As in the case $s=1$, the spans of the respective currents
\eqref{Tspan}, \eqref{Zspan}, \eqref{Vspan}
for fixed $r\geq 0$
contain non-trivial currents of lower weight,
and consequently it is convenient to introduce quotients of
the vector spaces $\vs{2r}{T}$, $\vs{2r+1}{Z}$, $\vs{2r+2}{V}$
of equivalence classes of these currents
of weight at most $2r$, $2r+1$, $2r+2$, respectively.
Let
\EQ
\qvs{2r}{T}=\vs{2r}{T}/\vs{2r-2}{T} ,\quad
\qvs{2r+1}{Z}=\vs{2r+1}{Z}/\vs{2r-1}{Z} ,\quad
\qvs{2r+2}{V}=\vs{2r+2}{V}/\vs{2r}{V} ,\quad \eqtext{ for $r\ge 1$, }
\endEQ
and
\EQ
\qvs{0}{T}=\vs{0}{T} ,\quad
\qvs{1}{Z}=\vs{1}{Z} ,\quad
\qvs{2}{V}=\vs{2}{V} .
\endEQ

\Proclaim{ Corollary 4.4. }{
The vector space of all quadratic currents of weight at most $w$
is isomorphic to the direct sum
\EQ\label{iso}
\oplus_{r=0}^{[w/2]} \qvs{2r}{T}
\oplus_{r=0}^{[(w-1)/2]} \qvs{2r+1}{Z}
\oplus_{r=0}^{[w/2-1]} \qvs{2r+2}{V}.
\endEQ
Moreover, for each $r\ge 0$,
equations \eqsref{highestorderTchar}{highestorderVchar}
establish a one-to-one correspondence
between the quotient spaces of quadratic currents
$\qvs{2r}{T}$, $\qvs{2r+1}{Z}$, $\qvs{2r+2}{V}$
and the vector spaces of \Kspin/s of types
$(2s+2r-1,2s+2r-1)_R$, $(2s+2r,2s+2r)_R$, $(2r+1,2r+4s+1)$,
respectively. }

We now outline the proof of the main classification Theorems~2.2 and~2.3
based on the preceding results.
First, by calculations similar to those in the proof of Lemma~4.2,
we find the following relation between the characteristics of
the quadratic currents
\eqref{Tspan}, \eqref{Zspan}, \eqref{Vspan}
and those of the quadratic currents
\eqref{Tdensity}, \eqref{Zdensity}, \eqref{Vdensity}
in Theorem~2.2.

\Proclaim{ Proposition 4.5. }{
Let $\prodKV{}{AA'}{}$ be a real conformal \Kvec/
and let 
$\prodKV{(k)}{A_1A'_1\cdots A_kA'_k}{}
= \prodKV{}{(A_1|A'_1|}{} \cdots \prodKV{}{A_k)A'_k}{}$
denote the rank $k$ conformal \Kten/ defined by
the $k$-fold symmetrized product of $\prodKV{}{AA'}{}$. 
Let $\prodKY{}{A'B'}{}$ be a self-dual conformal \KYten/
and let the rank $k$ conformal \KYten/ defined by
the $k$-fold symmetrized product of $\prodKY{}{}{A'B'}$ be denoted
$\prodKY{(k)}{A'_1\cdots A'_{4k}}{}
= \prodKY{}{(A'_1A'_2}{} \cdots \prodKY{}{A'_{4k-1}A'_{4k})}{}$.
Then the stress-energy currents
$\curr{(r)}{T}{A'}{A}(\phi,\bar\phi;\varrho)$,
zilch currents
$\curr{(r)}{Z}{A'}{A}(\phi,\bar\phi;\varrho)$,
and chiral currents
$\curr{(r)}{V}{A'}{A}(\phi,\bar\phi;Y,\varrho)$
admit the respective characteristics
\EQs
&&
(-1)^r 2\TQ{2r}{A'_1\cdots A'_{2s-1}}{A}[\varrho^{(2s-1)},\varrho] ,
\\
&&
(-1)^r 2\ZQ{2r+1}{A'_1\cdots A'_{2s-1}}{A}[\varrho^{(2s-1)},\varrho] ,
\\
&&
(-1)^{r+1} \sum_{p=0}^{r+1}
\VQ{2r+1}{A'_1\cdots A'_{2s-1}}{A}[(\Lie{\varrho})^pY^{(2s-1)},\varrho] .
\endEQs
These characteristics are adjoint symmetries of order
$q_T=2r$, $q_Z=2r+1$, $q_V=2r+2$,
whose highest order terms are given by
\eqref{highestorderTchar},
\eqref{highestorderZchar},
\eqref{highestorderVchar},
with $\zeta=\varrho$, $\xi=\varrho^{(2s-1)}$, $\Upsilon=Y^{(2s-1)}$. }

One can now use the \Kspin/ factorization in Lemma~3.1,
and Corollary~4.4 combined with Proposition~4.5,
to show by induction that
for each $r\ge 0$
the characteristics
\EQ
\TQ{2r}{A'_1\cdots A'_{2s-1}}{A}[\xi,\zeta], \quad
\ZQ{2r+1}{A'_1\cdots A'_{2s-1}}{A}[\xi,\zeta], \quad
\VQ{2r+1}{A'_1\cdots A'_{2s-1}}{A}[\Upsilon,\zeta]
\endEQ
given by \eqsref{Tchar}{Vchar}
are, respectively,
equivalent to a linear combination of the characteristics
\EQ
\TQ{2n}{A'_1\cdots A'_{2s-1}}{A}[\varrho^{(2s-1)},\varrho] , \quad
\ZQ{2n+1}{A'_1\cdots A'_{2s-1}}{A}[\varrho^{(2s-1)},\varrho] , \quad
\VQ{2n+1}{A'_1\cdots A'_{2s-1}}{A}[(\Lie{\varrho})^pY^{(2s-1)},\varrho]
\endEQ
involving real conformal \Kvec/s $\varrho$
and self-dual conformal \KYten/s $Y$
for $0\le n\le r$.
Hence, Theorem~2.5 now leads to the following classification result.

\Proclaim{ Theorem 4.6. }{
For each integer $r\ge 0$,
the currents
$\TH{2r}{AA'}{}[\xi,\zeta]$,
$\ZH{2r+1}{AA'}{}[\xi,\zeta]$,
$\VH{2r+1}{AA'}{}[\Upsilon,\zeta]$
are respectively equivalent to
a sum of
stress-energy currents
$\curr{(n)}{T}{A'A}{}(\phi,\bar\phi;\varrho)$,
zilch currents
$\curr{(n)}{Z}{A'A}{}(\phi,\bar\phi;\varrho)$,
and chiral currents
$\curr{(n)}{V}{A'A}{}(\phi,\bar\phi;Y,\varrho)$,
$0\le n\le r$,
involving real conformal \Kvec/s $\varrho$
and self-dual conformal \KYten/s $Y$,
for each $n$. }

Finally, from Lemma~3.1 and Corollary~4.4,
we are able to count the number of linearly independent
equivalence classes of conserved currents of any weight,
which establishes Theorem~2.3.

\Proclaim{ Corollary 4.7. }{
The respective dimensions of the vector spaces of quadratic currents
$\qvs{w}{Z}$ of odd weight $w\ge 1$ 
and $\qvs{w}{T} \oplus \qvs{w}{V}$ of even weight $w\ge 2$ 
are $(2s+w)^2 (2s+w+1)^2 (4s+2w+1)/12$
and $(4s+2w+1) ( 2w(w+1)(4s+w)(4s+w+1) +(2s+w)^2 (2s+w+1)^2 )/12$. }


\section{ Conserved quantities }
\label{quantities}

Let $\x{\mu}{}=\{ t,x^1,x^2,x^3 \}$
denote the standard Minkowski spacetime coordinates and
let $\Sigma_t$ denote a spacelike hyperplane $t=\const$.
Given a locally constructed conserved current $\dens{AA'}$
of the massless spin $s$ field equations \eqref{Feq},
let
\EQ
\Cquant{}{\dens{}[\phi]} = \int_{\Sigma_t} \t{}{AA'} \dens{AA'}[\phi] d^3x ,
\endEQ
where $\t{AA'}{}$ is the unit timelike future normal to $\Sigma_t$,
and $\dens{AA'}[\phi]$ denotes $\dens{AA'}$ evaluated on
a smooth solution $\sF{}{A_1\cdots A_{2s}}(t,x)$ of \eqref{Feq}.
Then $\Cquant{}{\dens{}[\phi]}$ is
a conserved (\ie/, time-independent) finite quantity,
$\Cquant{}{\dens{}[\phi]} <\infty$ and $\der{t}\Cquant{}{\dens{}[\phi]}=0$,
provided that
the density $\t{}{AA'} \dens{AA'}[\phi]$ decays sufficiently fast
and the flux of the spatial projection of $\dens{AA'}[\phi]$
vanishes at infinity on $\Sigma_t$.
In particular, this holds if $\sF{}{A_1\cdots A_{2s}}(t,x)$
is compactly supported on $\Sigma_t$ for all $t$,
and if $\dens{AA'}[\phi](t,x)=0$ whenever
$\sder{C'_1}{C_1}\cdots\sder{C'_p}{C_p}\sF{}{A_1\cdots A_{2s}}(t,x)=0$,
$p\ge 0$,
as is the case with the linear and quadratic currents
\eqsref{Wspan}{Vspan}.

We now give a derivation of conserved quantities
arising from the stress-energy, zilch, and chiral currents
\eqsref{Tspan}{Vspan}
based on a $3+1$ split of the spin $s$ field equations \eqref{Feq}
into electric and magnetic parts.

\subsection{ Electric and magnetic spin $s\ge 1/2$ fields }

As a preliminary, we consider the familiar case $s=1$.
Using a timelike unit vector $\t{AA'}{}$,
we define the electric and magnetic parts of
the $s=1$ field strength $\sF{}{AB}$ by
$\E{}{AA'}+\i \B{}{AA'} = \t{B'}{A} \csF{}{A'B'}$,
where $\E{}{AA'},\B{}{AA'}$ represent real-valued vectors
satisfying
$\t{AA'}{}\E{}{AA'} = \t{AA'}{}\B{}{AA'} =0$.
Thus $\E{}{AA'},\B{}{AA'}$ have no time component,
and we thereby obtain the decomposition
$\csF{}{A'B'} = 2\t{A}{A'}( \E{}{AB'} +\i \B{}{AB'} )$.
Now, we split the $s=1$ field equations
$\csFder{}{A'B',}{B'}{A}=0$
into time and space components with respect to $\t{AA'}{}$.
Let $\D{t} = \t{AA'}{}\D{AA'}$
denote the total time derivative,
and $\vecD{AA'}= \D{AA'}- \t{}{AA'}\D{t}$
denote the total spatial gradient.
The splitting then yields
\EQs
&&
\D{t} \E{}{AA'} = \vecD{} \times \B{}{AA'} ,\qquad
\vecD{} \cdot \E{}{} =0 ,
\label{Eeq}\\
&&
\D{t} \B{}{AA'} = -\vecD{} \times \E{}{AA'} ,\qquad
\vecD{} \cdot \B{}{} =0 ,
\label{Beq}
\endEQs
where $\vecD{} \cdot$ and $\vecD{} \times$
denote the standard spatial divergence and curl operators
which act on spinorial vector functions $\spinor{v}{}{AA'}$
satisfying $\t{AA'}{}\spinor{v}{}{AA'} =0$
on $J^\infty(\phi)$ by
\EQs
\vecD{}\cdot v = \veccoD{AA'} \spinor{v}{}{AA'} ,\qquad
\vecD{}\times \spinor{v}{}{AA'} =
\i \t{BB'}{}( \vecD{AB'} \spinor{v}{}{BA'} - \vecD{BA'} \spinor{v}{}{AB'} )
.
\endEQs
Thus, equations \eqrefs{Eeq}{Beq} describe
an electric-magnetic formulation of the $s=1$ field equations,
comprising a spinorial version of the Maxwell equations.

We now proceed analogously for integer spins $s=1,2,\ldots$.
Write
\EQ
\t{B'_1}{A_1} \cdots \t{B'_s}{A_s} \csF{}{A'_1\cdots A'_s B'_1\cdots B'_s}
= \E{}{A_1\cdots A_s A'_1\cdots A'_s}
+\i \B{}{A_1\cdots A_s A'_1\cdots A'_s} ,
\label{EBdef}
\endEQ
where
$\E{}{A_1\cdots A_s A'_1\cdots A'_s},\B{}{A_1\cdots A_s A'_1\cdots A'_s}$
are real symmetric spinors
satisfying
\EQ
\t{A_1A'_1}{} \E{}{A_1\cdots A_s A'_1\cdots A'_s}
= \t{A_1A'_1}{} \B{}{A_1\cdots A_s A'_1\cdots A'_s} =0 .
\endEQ
This expression \eqref{EBdef} yields the identity
\EQ
\csF{}{A'_1\cdots A'_s B'_1\cdots B'_s}
= 2^s \t{A_1}{A'_1} \cdots \t{A_s}{A'_s}
( \E{}{A_1\cdots A_s B'_1\cdots B'_s}
+\i \B{}{ A_1\cdots A_s B'_1\cdots B'_s} ) ,
\label{EBFid}
\endEQ
and hence the spin $s$ field equations \eqref{Feq}
split into the components
\EQs
&&
\D{t} \E{}{A_1\cdots A_s A'_1\cdots A'_s}
= \vecD{} \times \B{}{A_1\cdots A_s A'_1\cdots A'_s} ,\quad
\D{t} \B{}{A_1\cdots A_s A'_1\cdots A'_s}
= -\vecD{} \times \E{}{A_1\cdots A_s A'_1\cdots A'_s} ,
\label{scurleq}\\
&&
\vecD{} \cdot \E{}{A_1\cdots A_{s-1} A'_1\cdots A'_{s-1}}  =0 ,\quad
\vecD{} \cdot \B{}{A_1\cdots A_{s-1} A'_1\cdots A'_{s-1}} =0 .
\label{sdiveq}
\endEQs
Here the action of the operator $\vecD{}\times$
on symmetric spinors in \eqref{scurleq}
is well defined due to the divergence conditions \eqref{sdiveq}.

Note the field equations \eqrefs{scurleq}{sdiveq}
admit duality rotations generated by the transformation
on the electric and magnetic spinors
\EQ
\E{}{A_1\cdots A_s A'_1\cdots A'_s} \rightarrow
-\B{}{A_1\cdots A_s A'_1\cdots A'_s} ,\qquad
\B{}{A_1\cdots A_s A'_1\cdots A'_s} \rightarrow
\E{}{A_1\cdots A_s A'_1\cdots A'_s}
\label{EBdual}
\endEQ
corresponding to the symmetry \eqref{sduality}.
In tensor form,
these spinors represent electric and magnetic
trace-free symmetric real tensors
$\E{}{a_1\cdots a_s} = \e{a_1}{A_1A'_1}\cdots\e{a_s}{A_sA'_s}
\E{}{A_1\cdots A_s A'_1\cdots A'_s}$,
$\B{}{a_1\cdots a_s} = \e{a_1}{A_1A'_1}\cdots\e{a_s}{A_sA'_s}
\B{}{A_1\cdots A_s A'_1\cdots A'_s}$
(see also \Ref{Senovilla}).

One can generalize the previous equations to half-integer spins
$s=1/2,3/2,\ldots$
by considering a complex-valued hybrid trace-free symmetric tensor/spinor
$\E{}{Aa_1\cdots a_{s-1/2}} 
= \e{a_1}{A_1A'_1}\cdots\e{a_{s-1/2}}{A_{s-1/2} A'{}_{s-1/2}}
\E{}{AA_1\cdots A_{s-1/2} A'_1\cdots A'{}_{s-1/2}}$
determined by
\EQ
\E{}{A_1\cdots A_{s+1/2} A'_1\cdots A'{}_{s-1/2}}
= \t{B'_1}{A_1} \cdots \t{B'{}_{s+1/2}}{A_{s+1/2}}
\csF{}{A'_1\cdots A'{}_{s-1/2} B'_1\cdots B'{}_{s+1/2}} ,
\endEQ
for $s=j+1/2$, $j\ge 0$. 
Here the symmetric spinor $\E{}{A_1\cdots A_{s+1/2} A'_1\cdots A'{}_{s-1/2}}$
satisfies
\EQ
\csF{}{A'_1\cdots A'{}_{s-1/2} B'_1\cdots B'{}_{s+1/2}}
= 2^{s+1/2} \t{A_1}{B'_1} \cdots \t{A_{s+1/2}}{B'{}_{s+1/2}}
\E{}{A_1\cdots A_{s+1/2} A'_1\cdots A'{}_{s-1/2}}
\label{EBFid'}
\endEQ
and $\t{A_1A'_1}{} \E{}{A_1\cdots A_{s+1/2} A'_1\cdots A'{}_{s-1/2}} =0$,
but it has no well-defined decomposition into
real and imaginary (respectively electric and magnetic) parts.
Consequently, the time and space components of
the spin $s$ field equations \eqref{Feq}
now yield, after some algebraic manipulations,
\EQs
&&
\D{t} \E{}{A_1\cdots A_{s+1/2} A'_1\cdots A'{}_{s-1/2}}
= -\i\vecD{} \times \E{}{A_1\cdots A_{s+1/2} A'_1\cdots A'{}_{s-1/2}} , \\
&&
\vecD{} \cdot \E{}{A'_1\cdots A'{}_{s-3/2} A_1\cdots A_{s-1/2}} =0 ,\quad
\eqtext{ if $s=j+1/2$, $j\ge 1$, }\\
&&
\D{t} \E{}{A}
= -\i\vecD{+} \times \E{}{A} , \quad
\eqtext{ if $s=1/2$, }
\label{zeroscurleq}
\endEQs
where the curl operator in \eqref{zeroscurleq}
acts on spinor functions $\spinor{v}{}{A}$
on $J^\infty(\phi)$ by
$\vecD{+}\times \spinor{v}{}{A} =
2\i \t{BB'}{} \vecD{AB'} \spinor{v}{}{B}$.
Note these field equations admit duality rotations
generated by the transformation
\EQ
\E{}{A_1\cdots A_{s+1/2} A'_1\cdots A'{}_{s-1/2}} \rightarrow
\i\E{}{A_1\cdots A_{s+1/2} A'_1\cdots A'{}_{s-1/2}}
\quad\eqtext{ for $s=j+1/2$, $j\ge 0$. }
\label{Edual}
\endEQ

\subsection{ Electric and magnetic conserved quantities }

The stress-energy, zilch, and chiral currents
\eqref{Tcurr}, \eqref{Zcurr}, \eqref{Vcurr}
can be expressed straightforwardly
in electric-magnetic form \eqrefs{EBFid}{EBFid'}
for all $s\ge 1/2$.
Here we write down the resulting conserved quantities
obtained from the associated conserved tensors
\eqref{Ttens}, \eqref{Ztens}, \eqrefs{V+tens}{V-tens}
in some simple cases.

Consider integer spins $s=1,2,\ldots$.
We will use a dot notation to denote contraction of tensor indices.
A complete contraction of the spin $s$ energy tensor and zilch tensor
with the timelike vector $\t{AA'}{}$ yields
\EQs
\frac{1}{2} \t{A_1}{A'_1} \cdots \t{A_{2s}}{A'_{2s}}
\T{T}{A'_1\cdots A'_{2s}}{A_1\cdots A_{2s}}
&&
= \E{}{} \cdot \E{}{} +\B{}{} \cdot \B{}{} ,
\label{EBenergy}\\
\frac{1}{2} \t{A_1}{A'_1} \cdots \t{A_{2s}}{A'_{2s}} \t{B}{B'}
\T{Z}{A'_1\cdots A'_{2s}B'}{A_1\cdots A_{2s}B}
&&
= -\E{}{} \cdot( \vecD{} \times \E{}{} )
- \B{}{} \cdot( \vecD{} \times \B{}{} ) ,
\label{EBzilch}
\endEQs
after the elimination of time derivatives by means of the field equations
\eqref{scurleq}.
The analogous contraction of $\t{AA'}{}$
with the spin $s$ chiral tensors vanishes.
However, by contracting with $\t{AA'}{}$ and $\u{BB'}{}$
on these tensors,
where $\u{BB'}{}$ is any fixed spacelike unit vector,
we obtain the expressions
\EQs
&&
\t{A_1}{A'_1} \t{A_2}{A'_2} \t{B_1}{B'_1} \cdots \t{B_{2s}}{B'_{2s}}
\u{C_1}{C'_1} \cdots \u{C_{2s}}{C'_{2s}}
\T{V_+}{A'_1A'_2 B'_1C'_1\cdots B'_{2s}C'_{2s}}
{A_1A_2 B_1C_1\cdots B_{2s}C_{2s}}
=
\nonumber\\&&
-( \u{(s)}{} \cdot( \vecD{} \times \E{}{} ) )^2
+ ( \u{(s)}{} \cdot( \vecD{} \times \B{}{} ) )^2
+\frac{1}{2} ( \vecD{} (\u{(s)}{} \cdot \E{}{}) )^2
-\frac{1}{2} ( \vecD{} (\u{(s)}{} \cdot \B{}{}) )^2 ,
\label{EBchiral+}\\
&&
\t{A_1}{A'_1} \t{A_2}{A'_2} \t{B_1}{B'_1} \cdots \t{B_{2s}}{B'_{2s}}
\u{C_1}{C'_1} \cdots \u{C_{2s}}{C'_{2s}}
\T{V_-}{A'_1A'_2 B'_1C'_1\cdots B'_{2s}C'_{2s}}
{A_1A_2 B_1C_1\cdots B_{2s}C_{2s}}
= \nonumber\\&&
\u{(s)}{} \cdot( \vecD{} \times \E{}{} )
\u{(s)}{} \cdot( \vecD{} \times \B{}{} )
+ \vecD{} (\u{(s)}{} \cdot \E{}{}) \cdot
\vecD{} (\u{(s)}{} \cdot \B{}{}) ,
\label{EBchiral-}
\endEQs
where the last two terms in both \eqrefs{EBchiral+}{EBchiral-}
involve a contraction of gradients,
and where $\u{(s)}{}$ denotes $\u{A_1}{A'_1}\cdots\u{A_s}{A'_s}$.

Note that, under the duality symmetry \eqref{EBdual},
the conserved quantities in \eqrefs{EBenergy}{EBzilch}
display manifest invariance,
whereas the conserved quantities in \eqrefs{EBchiral+}{EBchiral-}
are chiral.
The quantity \eqref{EBenergy} has an obvious interpretation
as a nonnegative energy density of spin $s$ electric and magnetic fields.
Similarly, the quantity \eqref{EBzilch} can be interpreted
as a signed power density
where the sign is determined according to decomposing
the spin $s$ electric and magnetic fields into
positive and negative frequency components.
Indeed, for positive/negative frequency plane waves
\EQ
\E{}{A_1\cdots A_s A'_1\cdots A'_s} +\i\B{}{A_1\cdots A_s A'_1\cdots A'_s}
= f \z{}{A_1 A'_1} \cdots \z{}{A_s A'_s}
e^{\pm\i\omega (t-x^3)} 
\endEQ
propagating in the $x^3$ direction
with amplitude $f$ and frequency $\omega$
and polarization vector 
$\z{}{AA'} = \spinor{x^1}{}{AA'}+\i\spinor{x^2}{}{AA'}$
(where $\spinor{x^\mu}{}{AA'}$ denotes
a unit vector in the spatial direction $\x{\mu}{}$),
the quantities \eqrefs{EBenergy}{EBzilch} yield
$|f|^2$ and $\mp\omega|f|^2$.
The chiral quantities \eqrefs{EBchiral+}{EBchiral-}
for $\u{}{BB'} =\z{}{BB'}$ yield
$\omega^2( f^2 + \bar f^2)$ and $\i\omega^2( f^2 - \bar f^2)$.

Similar expressions to
\eqref{EBenergy}, \eqref{EBzilch}, \eqrefs{EBchiral+}{EBchiral-}
are obtained for half-integer spins
$s=1/2,3/2,\ldots$.
However, the interpretation of these quantities changes.
In particular,
the quantity arising from the complete contraction of the tensor
$\T{T}{A'_1\cdots A'_{2s}}{A_1\cdots A_{2s}}$
with $\t{AA'}{}$
no longer represents an energy density
but instead is a ``spinor particle density''
$\t{}{AA'} \cE{A'}{} \cdot \E{A}{}$,
which is a spin $s$ generalization of 
the familiar neutrino (\ie/, $s=1/2$) particle density expression
$\t{AA'}{} \sF{}{A} \csF{}{A'} = 
2 \t{}{AA'} \cE{A'}{} \E{A}{}$ \cite{Wald}.

\subsection{ A complete set }

Here we outline an algorithm for constructing
an explicit basis for the vector spaces spanned by
the stress-energy, zilch, and chiral currents.

To begin we write explicit expressions for
conformal \Kten/s and conformal \KYten/s,
given by solutions of the \Kspin/ equations \eqref{KSeq}.
In spinor form,
complex-valued conformal \Kvec/s and self-dual conformal \KYten/s
are quadratic polynomials
\EQs
&& \KV{AA'}{} =
\spinor{\alpha_1}{AA'}{} + \alpha_2\x{AA'}{}
+ \spinor{\alpha_3}{A'B'}{}\x{A}{B'}
+ \spinor{\alpha_4}{AB}{}\x{A'}{B}
+\spinor{\alpha_5}{B'B}{}\x{A}{B}\x{A'}{B'},
\\
&& \KY{A'B'}{} =
\spinor{\beta_1}{A'B'}{} + \spinor{\beta_2}{A(A'}{}\x{B')}{A}
+\spinor{\beta_3}{AB}{}\x{A'}{A}\x{B'}{B},
\label{KS}
\endEQs
in the Minkowski spacetime coordinates $\x{CC'}{}$,
where
$\spinor{\alpha_1}{AA'}{}$,
$\spinor{\alpha_2}{}{}$,
$\spinor{\alpha_3}{A'B'}{}$,
$\spinor{\alpha_4}{AB}{}$,
$\spinor{\alpha_5}{BB'}{}$,
$\spinor{\beta_1}{A'B'}{}$,
$\spinor{\beta_2}{AA'}{}$,
$\spinor{\beta_3}{AB}{}$
are constant symmetric spinors.
Similarly,
a \Kspin/
$\KS{A_1\cdots A_k A'_1\cdots A'_{k+2l}}{}$
of type $(k,k+2l)$ is a polynomial of degree at most $2(k+l)$
in $\x{CC'}{}$
whose monomial terms of degree $0\le p\le 2(k+l)$ are given by
\EQ
\KS{\hp{(p)}A'_1\cdots \cdots A'_{k+2l}}{(p)A_1\cdots A_k}(x)
= \spinor{\gamma_{(p;q)}}
{B'_1\cdots B'_q (A'_{p-q+1} \cdots A'_{k+2l}}
{B_1\cdots B_{p-q} (A_{q+1} \cdots A_k}
\x{A'_1|B_1|}{} \cdots \x{A'_{p-q})B_{p-q}}{}
\x{}{A_1|B'_1|} \cdots \x{}{A_q)B'_q}
\label{KSmonomial}
\endEQ
with constant symmetric spinor coefficients
$\spinor{\gamma_{(p;q)}}{}{}$,
where
$0\le q\le \min(k,p)$.

Fix a spinor basis $\{\so{A}{},\si{A}{}\}$
satisfying $\so{}{A}\si{A}{}=1$.
This determines a corresponding null-tetrad basis for
the vector space of constant spinorial vectors
\EQ
\spinor{k}{AA'}{} = \so{A}{}\cso{A'}{} ,\quad
\spinor{\ell}{AA'}{} = \si{A}{}\csi{A'}{} ,\quad
\spinor{m}{AA'}{} = \so{A}{}\csi{A'}{} ,\quad
\cspinor{m}{AA'}{} = \si{A}{}\cso{A'}{} ,
\label{tetradbasis}
\endEQ
and of constant self-dual spinorial skew-tensors
\EQ
\cspinor{m}{A'A}{}\spinor{k}{\hp{A}B'}{A} = \cso{A'}{}\cso{B'}{} ,\quad
\spinor{\ell}{A(A'}{}\spinor{k}{\hp{A}B')}{A} = \cso{(A'}{}\csi{B')}{} ,\quad
\spinor{\ell}{AA'}{}\spinor{m}{\hp{A}B'}{A} = \csi{A'}{}\csi{B'}{} .
\label{skewtensbasis}
\endEQ
A basis for the monomial \Kspin/s \eqref{KSmonomial} is generated by
the set of constant spinors
\EQs
&&
\spinor{\gamma_{(p;q)}}
{C'_1\cdots C'_Q}{C_1\cdots C_R}
= \cso{C'_1}{} \cdots  \cso{C'_n}{} \csi{C'_{n+1}}{} \cdots \csi{C'_Q}{}
\so{}{C_1} \cdots  \so{}{C_m} \si{}{C_{m+1}} \cdots \si{}{C_R} ,
\label{KSbasis}\\&&
0\le n\le Q=k+2l+2q-p ,\quad
0\le m\le R=k+p-2q .
\nonumber
\endEQs

Now, recall from Corollary~4.4 that
the equivalence classes of currents of minimal even weight 
$w\ge 0$ in $\qvs{w}{T}$ and $w\ge 2$ in $\qvs{w}{V}$, 
and of minimal odd weight $w\ge 1$ in $\qvs{w}{Z}$
are in one-to-one correspondence with \Kspin/s respectively
of type $(w+2s-1,w+2s-1)_R$ and $(w-1,w+4s-1)$
if $w$ is even
and of type $(w+2s-1,w+2s-1)_R$
if $w$ is odd.
These \Kspin/s are given by
sums of monomials \eqref{KSmonomial} of degree
$0\le p\le 2w+4s-2$,
with $k=w+2s-1$, $l=0$, and, $k=w-1$, $l=2s$ when $w$ is even,
and $k=l=w+2s-1$ when $w$ is odd.
The algorithm now proceeds in three steps:
\vskip0pt
(i) given $w\geq 0$, then for each value of $p,q,n,m$,
choose a factorization of the monomial \Kspin/ basis
\eqrefs{KSmonomial}{KSbasis}
into a symmetric
product of complex conformal \Kvec/s and self-dual conformal \KYten/s;
\vskip0pt
(ii) substitute the null tetrad basis
\eqref{tetradbasis}, \eqref{skewtensbasis}
for pairs of basis spinors in the factorized monomial \Kspin/;
\vskip0pt
(iii) write the real part of the conserved currents
\eqref{Tcurr}, \eqref{Zcurr}, \eqref{Vcurr}
using the resulting conformal \Kvec/s and conformal \KYten/s.

This algorithm yields a basis for the quotient spaces of
stress-energy currents $\qvs{w}{T}$,
zilch currents $\qvs{w}{Z}$,
and chiral $\qvs{w}{V}$,
respectively for $w=2r,2r+1,2r+2$, 
for each fixed $r \geq0$.
Details of this construction in the case $s=1$ are given in \Ref{maxwell}.

\section{Concluding Remarks}
\label{remarks}

In this paper
we have presented a complete and explicit classification of
all locally constructed conserved currents
for massless linear symmetric spinor fields of spin $s\ge 1/2$
in Minkowski spacetime.
This work generalizes the recent classification results
we obtained \cite{maxwell}
for all conserved currents locally constructed from
the electromagnetic field strength in the case $s=1$.
Our results give a spin $s\ge 1/2$ generalization of
the new electromagnetic chiral tensor found in the spin $1$ case,
in addition to spin $s\ge 1/2$ generalizations of
Lipkin's electromagnetic zilch tensor
\cite{Lipkin,Kibble}
and the well-known electromagnetic energy tensor
\cite{Bessel-Hagen}.
The chiral tensor is physically interesting as it possesses
odd parity under the interchange of the electric and magnetic
parts of the spin $s=1,2,\ldots$ field strengths
(and under a phase rotation on the spin $s=1/2,3/2,\ldots$ field strengths),
in contrast to the even parity of both the energy and zilch tensors.
In particular,
the duality symmetry of the spin $s$ field equations is broken by
the conserved currents associated with the chiral tensor,
and hence these chiral currents
distinguish between pure ``electric'' and pure ``magnetic'' field strengths.

Moreover, our results yield a complete set of conserved quantities
locally constructed from the spin $s\ge 1/2$ field strength.
While the physical interpretation of these quantities apart from
the well-known energy, momentum, stress, angular and boost momentum
obtained from the energy tensor
have yet to be fully explored,
they provide new constants of motion characterizing
the propagation of massless spin $s\ge 1/2$ fields in flat spacetime,
which is of obvious interest in the study of, \eg/,
gravitons ($s=2$), neutrinos ($s=1/2$), and gravitinos ($s=3/2$).
In particular,
the zilch tensor yields physically meaningful quantities
related to the positive/negative frequency power spectrum of
the propagating fields.

Our classification results are also applicable
to the problem of constructing all consistent nonlinear
interactions of spin $s\ge 1/2$ gauge fields.
As shown in the work in \Refs{Anco,Henneaux1,Henneaux2},
conserved currents of the linear field equations
play a central role in the construction
by determining possible quadratic interaction terms for
nonlinear spin $s$ gauge field equations.

Finally, our methods and results can be extended directly
from the linear massless spin $s\ge 1/2$ field equations in flat spacetime
to the corresponding equations
\EQ
\covsder{A_1}{A'} \sF{}{A_1\cdots A_{2s}}(x)=0
\label{curvedeq}
\endEQ
in any locally conformally flat spacetime.
Here $\covsder{}{AA'} =\inve{a}{AA'} \covder{a}$
is the spinorial covariant derivative compatible with
the curved spacetime metric
$\g{ab} =\e{a}{AA'} \e{b}{BB'} \vol{AB}{}\vol{A'B'}{}$,
where $\e{a}{AA'}$ is a soldering form.
All locally constructed conserved currents
continue to arise from adjoint symmetries of \eqref{curvedeq}
through the integral formula \eqref{currformula}.
The adjoint symmetries can be obtained
by solving
\EQ
\sD{A(A'_1}{} \P{A'_2\cdots A'_{2s})}{A} =0
\endEQ
on the solution space of \eqref{curvedeq},
where $\P{A'_1\cdots A'_{2s-1}}{A}$
is a function of the spacetime coordinates
and the spin $s\ge 1/2$ field strength
and its symmetrized covariant derivatives,
and where $\sD{A}{A'}$ now denotes the covariant total derivative operator.
Up to equivalence,
$\P{A'_1\cdots A'_{2s-1}}{A}$ remains linear in the field variables,
with the coefficients determined by conformally-flat spacetime \Kspin/s
which satisfy the equations
\EQ
\covsder{(B'|(B}{} \KS{A_1 \cdots A_k) |A'_1 \cdots A'_l)}{} =0 .
\label{curvedKSeq}
\endEQ
The invariance of \eqsref{curvedeq}{curvedKSeq}
under local conformal scalings of $\g{ab}$ allows for
a complete and explicit classification of
the resulting spin $s$ conserved currents.
Corresponding conserved tensors arise from the conserved currents,
as in flat spacetime.

Note that the restriction to conformally flat metrics is necessary
for the well-posedness (local solvability)
of the field equations \eqref{curvedeq} for $s>1$,
due to the well-known algebraic consistency conditions \cite{Penrose,Wald}
occurring on solutions of \eqref{curvedeq}.
In the cases $s=1/2$ and $s=1$, the field equations \eqref{curvedeq}
are well-posed (locally solvable) without conditions on $\g{ab}$.
However,
the \Kspin/ equations \eqref{curvedKSeq}
possess, independently of $s$, similar algebraic consistency conditions
relating the \Kspin/ and the curvature spinor of $\g{ab}$ \cite{Penrose}.
Consequently, non-trivial spin $1/2$ and spin $1$ conserved currents
exist only for certain curved metrics $\g{ab}$,
as pointed out in \Ref{maxwell}.
Particularly interesting here is the Kerr blackhole spacetime metric,
for which the geodesic equations are known to possess
an extra conserved quantity related to the existence of a \KYten/
\cite{WalkerPenrose}.

Finally, in the spin $2$ case, recall that
the massless field strength $\sF{}{ABCD}$
describes \cite{Penrose,Wald}
perturbations of the spacetime metric $\g{ab}$
satisfying the vacuum Einstein field equations.
In particular,
if $\g{ab}$ is linearized around a background Minkowski metric $\flat{ab}$,
then the corresponding linearization of
the curvature spinor $\spinor{C}{}{ABCD}$ of $\g{ab}$
is given by $\sF{}{ABCD}$ satisfying \eqref{Feq}.
We thereby find that the spin $2$ energy tensor \eqref{Ttens}
corresponds to the spinorial form of the Bel-Robinson tensor
\cite{Penrose},
\EQ
\spinor{C}{}{ABCD} \cspinor{C}{A'B'C'D'}{} .
\endEQ
By the same correspondence,
the spin $2$ zilch tensor \eqref{Ztens}
and chiral tensors \eqrefs{V+tens}{V-tens}
lead to analogous tensors constructed from derivatives of
the curvature spinor, namely,
\EQ
\i( \cspinor{C}{A'B'C'D'}{} \covsder{E'}{E} \spinor{C}{}{ABCD}
- \spinor{C}{}{ABCD} \covsder{E'}{E} \cspinor{C}{A'B'C'D'}{} ) ,
\label{GRzilch}
\endEQ
and
\EQs
&&
\vol{AB}{}\vol{CD}{}\vol{EF}{}\vol{GH}{}
( \covsder{M'}{(M} \cspinor{C}{A'B'C'D'}{}
\covsder{N'}{N)} \cspinor{C}{E'F'G'H'}{}
-\covsder{N'}{[N} \cspinor{C}{A'B'C'D'}{}
\covsder{M'}{M]} \cspinor{C}{E'F'G'H'}{} ) + c.c.,
\label{GRchiral+}
\\
&&
\i\vol{AB}{}\vol{CD}{}\vol{EF}{}\vol{GH}{}
( \covsder{M'}{(M} \cspinor{C}{A'B'C'D'}{}
\covsder{N'}{N)} \cspinor{C}{E'F'G'H'}{}
-\covsder{N'}{[N} \cspinor{C}{A'B'C'D'}{}
\covsder{M'}{M]} \cspinor{C}{E'F'G'H'}{} ) + c.c.,
\label{GRchiral-}
\endEQs
where $c.c.$ denotes the complex conjugates of all preceding terms.
While the gravitational zilch tensor \eqref{GRzilch} 
has long been known \cite{zilch1,zilch2} 
as a counterpart of the electromagnetic zilch tensor \cite{Lipkin,Kibble},
the gravitational chiral tensors \eqrefs{GRchiral+}{GRchiral-} 
are, apparently, new. 
Discussion of the properties and significance of these chiral tensors
in General Relativity will be left for elsewhere.

\end{document}